\documentclass{aa}  
\usepackage{graphicx}
\usepackage{placeins}
\usepackage{siunitx}
\usepackage{multirow}
\usepackage[normalem]{ulem}
\usepackage{float}
\usepackage{comment}
\usepackage{capt-of}
\usepackage{txfonts}
\usepackage[colorlinks=true, citecolor=blue]{hyperref}
\usepackage{diagbox}
\usepackage{lastpage}
    
\begin{document} 
   \title{Impact of stellar population models on the estimated physical properties of galaxies}

   \author{Bozhidara Stoyanova
          \inst{1,2}
          \and
          Médéric Boquien\inst{2}
          \and
          Paola Santini\inst{3}
          \and
          Francesco Tombesi\inst{1,3,4,5}
          \and
          Emiliano Merlin\inst{3}
          \and
          Pietro Bergamini\inst{6}
          \and
          Véronique Buat\inst{7}
          \and
          Denis Burgarella\inst{7}
          }

   \institute{Università degli Studi di Roma Tor Vergata, Via Cracovia, 90, 00133 Roma RM, Italy
              \and Université Côte d'Azur, Observatoire de la Côte d'Azur, CNRS, Laboratoire Lagrange, F-06000 Nice, France
              \and INAF --- Osservatorio Astronomico di Roma, via di Frascati 33, I-00078 Monte Porzio Catone, Italy
              \and Physics Department, Tor Vergata University of Rome, Via della Ricerca Scientifica 1, 00133 Rome, Italy
              \and INFN --- Rome Tor Vergata, Via della Ricerca Scientifica 1, 00133 Rome, Italy 
              \and INAF – OAS, Osservatorio di Astrofisica e Scienza dello Spazio di Bologna, via Gobetti 93/3, I-40129 Bologna, Italy
              \and Aix Marseille Univ, CNRS, CNES, LAM, Marseille, France
             }

   \date{}
 
  \abstract
   {Accurate estimates of fundamental physical properties of galaxies, such as star formation rates (SFRs) or stellar masses, are essential for testing and constraining models of galaxy formation and evolution. Spectral energy distribution (SED) modeling has become the standard method for deriving these quantities. However, the influence of the underlying stellar population synthesis (SPS) models on the inferred parameters remains poorly quantified.}
   {This work investigates how the choice of SPS models affects the estimation of SFRs and stellar masses derived from SED modeling.}
   {Four widely used SPS models were applied to a sample of 17,230 galaxies with spectroscopic redshifts, selected from recently published Hubble Space Telescope and James Webb Space Telescope photometric catalogs. SEDs were modeled using the Code for Investigating GALaxy Emission. The analysis was performed in two steps: (i) estimating galaxy properties with each SPS model, and (ii) employing synthetic catalogs to assess the relative impact of model choice on the recovered parameters.}
   {Systematic differences are found among the models, with stellar mass estimates varying by up to $\sim0.6$~dex and SFRs by up to $\sim0.4$~dex between certain model pairs. The choice of stellar population model introduces significant systematic uncertainties in derived galaxy properties. This dependence should be accounted for when interpreting SED-based measurements and comparing results across different studies of galaxy evolution.}
  {}

   \maketitle
\section{Introduction} \label{sec:intro}
Peering into galaxies across the universe not only provides us with a glimpse into a distant past, but it also sheds light on the fundamental processes that drive galaxy formation and evolution. In this regard, measuring the physical properties of large samples of galaxies over a broad range of redshifts is key for constraining galaxy evolution models \citep{conselice2014a, madau2014a, somerville2015a}. 

The launch of the James Webb Space Telescope \citep[JWST,][]{gardner2023a} has marked a major advance in the study of the formation and evolution of galaxies. Thanks to its unparalleled sensitivity and spatial resolution, JWST can unveil previously unobserved populations of distant galaxies while providing us with exceptional data both in quality and wavelength coverage \citep{rigby2023a}. In its first years of operation, JWST has already discovered spectroscopically confirmed galaxies at redshifts up to $z \sim 14$ \citep{naidu2022a, curtislake2023a, finkelstein2023a, robertson2023a, robertson2024a}. The richness of the dataset at our disposal now allows us to explore galaxies both individually and as a population over a broad range of redshifts.

A key challenge is now to convert this treasure trove of observations to physical quantities \citep{laigle2019a}. In principle, spectroscopic observations are ideal for constructing a detailed and comprehensive picture of a galaxy. The combination of emission and absorption lines provides information on the dust and metal content, stellar populations, dynamics, and redshift \citep[e.g.,][]{kewley2019a, maiolino2019a}. However, obtaining spectroscopic data for large samples is both expensive and time-consuming. As a result, the largest extragalactic surveys remain predominantly photometric \citep[e.g.,][]{york2000a, cenarro2019a, ivezic2019a, casey2023a, mellier2025a}, with only a small fraction of galaxies observed spectroscopically, even with large dedicated surveys \citep{lefevre2015a, mclure2018a, desi2024a}. This is also the case for current photometric and spectroscopic JWST surveys \citep{casey2023a, eisenstein2023a, finkelstein2025a}. In consequence, robust methods for extracting the physical properties of galaxies from photometric data remain essential for advancing our understanding of galaxy evolution.

One of the leading techniques to measure the physical properties of galaxies is the forward-modeling of their spectral energy distribution \citep[SED,][]{walcher2011a}. In essence, SED modeling constructs the spectral emission of galaxies, including complex stellar populations, ionized gas, dust in absorption and in emission, and active galactic nuclei, all the while taking into account the effect of the intervening neutral gas in the intergalactic medium (IGM) to finally compute the corresponding fluxes in photometric bands \citep{conroy2013a}. The physical properties can then be statistically determined by fitting a set of such models to the observations. Ultimately, this method allows for the modeling of large samples of galaxies at a modest cost, both observationally and computationally.

One of the major hurdles we are facing is that the modeling assumptions that can be deeply ingrained in each physical component and in different modeling codes can lead to differences in the estimate of galaxy properties \citep{pacifici2023a, osborne2024a, bellstedt2025a}. Some of these assumptions are embedded in the single stellar population (SSP) models, which are a key component of SED models targeting UV-to-near-infrared wavelengths \citep{bruzual2003a, maraston2005a, conroy2009a, eldridge2009a, conroy2010b}. They describe the evolution with time of an instantaneous episode of star formation with a given initial mass function (IMF), which is generally assumed to be fully sampled, and at a certain metallicity. These models can then be used to compute the stellar spectrum of a galaxy with a complex star formation history by combining the contributions of multiple SSPs that have progressively assembled across cosmic time \citep{conroy2013a}. Different SSP models come with different sets of assumptions: stellar tracks and atmospheres, mass loss, IMF, metallicity, among others \citep[e.g.,][]{conroy2009a}. While different models commonly provide SSPs at different IMFs and metallicities, other parameters are often intrinsic to each model. This is the case, for instance, for the inclusion of binary evolution \citep{eldridge2009a, eldridge2017a, stanway2018a} or stellar rotation \citep{levesque2012a, leitherer2014a}, which can have a dramatic effect on the UV and ionizing spectrum in particular. These and other assumptions ultimately lead to differences in the predicted spectra for stellar populations that otherwise have the same age, IMF, and metallicity \citep{cidfernandes2014a, ge2019a}. 

Even though in recent years some studies have started to tackle the question of the effect of modeling choices on the measurement of physical properties \citep{pacifici2023a, osborne2024a, bellstedt2025a}, the impact of the choice of different SSP models on the measurement of the physical properties of galaxies across cosmic time is still not well understood. This question is increasingly pressing as different models are now available and routinely used, though not necessarily providing fully consistent results. Considering the growing samples of distant galaxies observed with JWST, it appears timely to assess the consistency of the estimates of the properties of stellar populations using some of the most widely used SSP models in the literature.

In this study, we focus specifically on the effect of SSP models on two fundamental quantities to constrain galaxy evolution models: the star formation rate (SFR) and the stellar mass ($M_\star$). The SFR provides insights into the current pace at which the gas reservoir is transformed into stars, and the stellar mass provides an estimate of the integral of past star formation, with a dependence on the IMF and the exact star formation history (SFH). In the case of star-forming galaxies, SFR and $M_\star$ follow a fairly tight relationship, the so-called star-forming main sequence (SFMS), which has been found over a wide range of redshifts and has become one of the key metrics to constrain galaxy evolution models \citep{noeske2007a, elbaz2007a, daddi2007a, speagle2014a, santini2017a, popesso2023a, clarke2024a, cole2025a, rinaldi2025a}. It is therefore important to have a detailed understanding of whether our estimates of the SFR and $M_\star$ might have systematic biases stemming from the choice of specific SSP models.

To quantify the importance of SSP models on the determination of physical properties, we propose a two-pronged experiment. First, we compare the physical properties of galaxies observed with JWST, inferred from SED modeling using four different SSP models that are regularly used in the literature. Offsets between these estimates inform us about the typical differences one can expect when using one model rather than another at various redshifts. Because this does not provide any indication of the correctness of the estimates in an absolute sense, in a second step, we generate synthetic catalogs simulating galaxies with each of the four SSPs. Knowing the physical properties of the simulated objects by construction, these catalogs serve as a ground truth. We then fit these catalogs with each of the four SSPs to assess whether it is possible to retrieve the ground truth SFR and $M_\star$ within a reasonable margin.

This paper is structured as follows. We introduce the selected dataset of galaxies observed with the Hubble Space Telescope (HST) and JWST in Sect.~\ref{sec:data}. We present the SED modeling along with the different SSPs we have adopted in Sect.~\ref{sec:modeling}. We compare the inferred SFR and $M_\star$ in Sect.~\ref{sec:results} and discuss the results in Sect.~\ref{sec:discussion}, including the induced changes on the derived SFMS, before concluding in Sect.~\ref{sec:conclusion}. Throughout this article, we assume a \cite{planck2020a} cosmology for determining the luminosity distance of the targets from their redshift.

\section{Sample selection and data}
\label{sec:data}
We need to assess how different SSP models affect the determination of SFR and $M_\star$ in galaxies. To this end, we selected galaxies observed both with JWST and HST at near-infrared and optical wavelengths. Several deep extragalactic fields satisfy these criteria. In this work, we adopted the ASTRODEEP-JWST \citep{merlin2024a} sample. The dataset includes composite photometric catalogs based on data from eight deep-sky JWST observational programs: CEERS \citep[ERS 1345, PI Finkelstein,][]{finkelstein2022}, DDT2756 (PI Chen), GLASS-JWST \citep[ERS 1324, PI Treu,][]{treu2022}, GO3990 (PI Morishita), JADES \citep[GTO 1180, PI Eisenstein, and GTO 1210, PI Luetzgendorf,][]{eisenstein2023a}, NGDEEP \citep[PI Finkelstein,][]{bagley2024}, PRIMER (GO 1837, PI Dunlop), and UNCOVER \citep[GO 2561, PI Labbé,][]{bezanson2024}. It includes data in 16 bands: eight from HST (ACS F435W, F606W, F775W, and F814W; HST WFC3 F105W, F125W, F140W, and F160W) and eight from JWST (NIRCam F090W, F115W, F150W, F200W, F277W, F356W, F410M, and F444W).

These broad pass-bands span a total observed-frame wavelength range from \SI{0.44}{\micro\metre} to \SI{4.44}{\micro\metre}. This coverage ensures observations are sensitive to rest-frame emission of young and old stellar populations over a fairly wide range of redshifts. At $z \simeq 0.5$, the combination of HST and JWST NIRCam bands probes the full range of the optical to the near-infrared. From $z=1$, the observations are sensitive from the far-UV to the near-infrared. However, from $z=4$, they do not extend beyond the red end of the optical, and above $z=10$, even JWST bands only probe wavelengths no longer than the blue in the optical. While the downside of having such a broad range of redshifts is that it makes the sample heterogeneous, an important advantage for such a study is that it allows us to explore whether there is any systematic difference in the physical properties for different stellar population models that depend on the redshift.

One of the seven fields in the catalogs (Abell 2744) is a lensed field, so the sources in it need to be delensed for an accurate estimate of their physical properties. However, we only compare the effect of SED fitting assumptions. A delensing procedure would affect a source in the same way regardless of the stellar population model used; therefore, for our comparisons, we did not delens the galaxies in this field. When we look at the SFMS in Section~\ref{sec:discussion}, we use the amplification factors calculated from the model of \citet{bergamini2023a} to obtain the final SFR and $M_\star$ for the sources in this field.

Overall, the dataset contains approximately $530\times10^3$ sources. To avoid any bias or uncertainty originating from errors on the redshift, we selected only sources with measured spectroscopic redshifts (20,853 galaxies in total)
as provided by the ASTRODEEP-JWST catalog \citep[see in particular Sect. 5 of ][for a description of the spectroscopic redshifts]{merlin2024a}. In addition, we removed sources flagged as having problematic photometry measurements (less than $3\sigma$ detection, close to a border, flagged after visual inspection, etc.), sources with measurements in less than four JWST bands, known AGN \citep{merlin2024a}, little red dots \citep{Barro2024, kokorev2024, kocevski2024, perez-gonzalez2024, labbe2025}, and brown dwarfs \citep{hainline2024, holwerda2024}. This results in a sample of 17,980 sources across all seven fields. 

\begin{figure}
    \centering
    \includegraphics[width=\columnwidth]{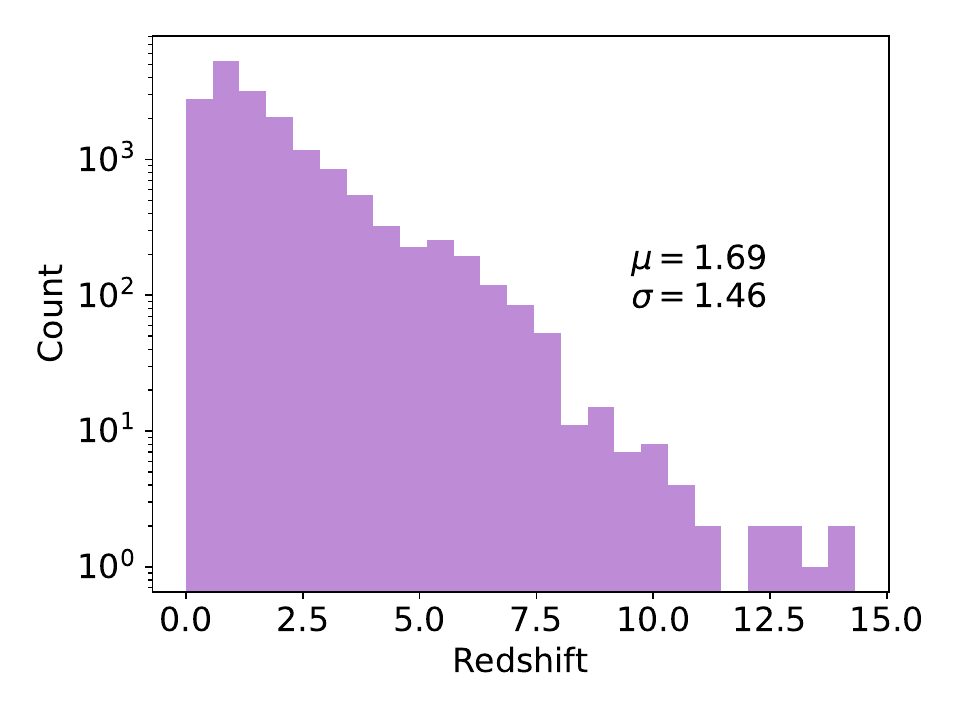}
    \caption{Redshift distribution of the final sample based on the ASTRODEEP-JWST catalog. The vast majority of sources are at low to moderate redshifts. Only a very small number of objects are beyond $z=8$.
    }
    \label{fig:z_distr}
\end{figure}

A final cut to the sample was also made after SED fitting, as a small number of sources could not be fit by some of the models. We removed these sources from all CIGALE runs, so our final sample is consistent between all models for both the observed and the synthetic galaxies. This final sample consists of 17,230 galaxies, and its redshift distribution is shown in Figure~\ref{fig:z_distr}.

\section{Spectral energy distribution modeling \label{sec:modeling}}

As mentioned earlier, SED modeling is a key technique for deriving the physical properties of galaxies from their multi-wavelength observations. In this work, we employed the Code Investigating GALaxy Emission \citep[CIGALE,][]{boquien2019a}, which we selected because it natively includes several of the most commonly used SSP models in galaxy evolution studies (see Sect.~\ref{ssec:SSP}) and is widely employed in the literature to characterize galaxies across cosmic time.

CIGALE constructs model spectra through a series of modules that sequentially describe the SFH, stellar populations, ionized gas (lines and continuum), dust attenuation, and the effect of the IGM and redshifting. Synthetic fluxes are then computed by integrating the model spectra within the 16 filter transmission curves of the ASTRODEEP-JWST catalog.

To infer the physical properties, CIGALE employs a Bayesian-like approach: grids of models are fitted to the observations, and the physical parameters and their associated uncertainties are derived from the likelihood-weighted means and standard deviations, respectively.

The detailed configuration of the modules and model grids is provided in Sect.~\ref{ssec:modeling}. Our goal here is not to explore in depth the influence of individual model parameters, but rather to compare the impact of adopting different SSP models while keeping the rest of the setup identical. For broader investigations of the effect of varying other model assumptions, we refer, for instance, to \cite{osborne2024a, bellstedt2025a}.

\subsection{Single stellar population models}
\label{ssec:SSP}
The central goal of this work is to assess how different SSP models influence the inferred SFR and $M_\star$. To this end, we considered four widely used stellar population models:
\begin{itemize}
    \item The \cite{bruzual2003a} models, often regarded as a reference or baseline due to their extensive use in the literature (Sect.~\ref{ssec:BC03}).
    \item The Charlot and Bruzual (2019) stellar populations, an updated version of \cite{bruzual2003a} incorporating major improvements and increasingly adopted in recent studies (Sect.~\ref{ssec:CB19}).
    \item The Binary population and spectral synthesis code (BPASS) models \citep[][Sect.~\ref{ssec:BPASS}]{eldridge2009a, eldridge2017a, stanway2018a}, which, unlike the previous models, include binary stellar evolution. We employed both the single-star and binary-evolution variants for comparison.
\end{itemize}

\subsubsection{Bruzual and Charlot (2003)}
\label{ssec:BC03}
The \cite{bruzual2003a} models are among the most widely used SSP models in extragalactic astronomy. They provide stellar population spectra covering ages from $10^5$~yr to $2 \times 10^{10}$~yr, with a high spectral resolution between \SI{320}{nm} and \SI{950}{nm}, and lower spectral resolution over a broader wavelength range from \SI{9.1}{nm} to \SI{160}{\micro\metre}. The models span 6 metallicities from $Z=0.0001$ to $Z=0.05$ and offer both the \cite{salpeter1955a} and \cite{chabrier2003a} IMF. This wide parameter space makes them highly versatile for modeling diverse galaxy populations across cosmic time.
 
\subsubsection{Charlot and Bruzual (2019)}
\label{ssec:CB19}
The so-called Charlot and Bruzual (2019) models are an evolution of the \cite{bruzual2003a} models \citep{gutkin2016a, plat2019a, sanchez2022a}. They feature updated treatments for hot massive stars and Asymptotic Giant Branch (AGB) stars, together with a much finer metallicity sampling of 15 values from $Z=0.0001$ to $Z=0.06$. \cite{plat2019a} showed that these single-star models reproduce properties slightly better than the BPASS v2.2.1 binary-star models (see Sect.~\ref{ssec:BPASS}), particularly in their prediction of the He~II ionizing photon production.

\subsubsection{Binary population and spectral synthesis}
\label{ssec:BPASS}

The BPASS models are distinguished by their explicit treatment of binary stellar evolution, incorporating the effects of mass transfer and interaction between binary companions. Such processes generally produce stellar populations that are bluer than those predicted by single-star models and emit larger numbers of Lyman continuum photons.

Although not all stars undergo binary interactions, these effects are particularly relevant because most massive stars are thought to experience mass exchange with a companion. While massive stars are rare, they dominate the energy output of star-forming galaxies. The precise fraction may vary with the stellar environment, but numerous studies have demonstrated that binary interactions significantly affect stellar populations across a wide range of environments and metallicities \cite[][and references therein]{eldridge2017a}. 

In this work, we use the BPASS v2.2 models \citep{stanway2018a}, which include several improvements over earlier versions, particularly for populations older than 1~Gyr. The updates comprise an expanded grid of low-mass stellar models and revised treatments of rejuvenation, dredge-up processes, and the AGB phase.

\subsection{Choice of modules for building galaxy SED models}
\label{ssec:modeling}

The available observations are sensitive to the combined effect of complex stellar populations, ionized gas, and dust attenuation. Since our selection excluded sources dominated by AGN and the filter set is not sensitive to dust emission, these components were not modeled. We therefore constructed our models using the following set of CIGALE modules.

\subsubsection{Star formation history}

The adopted SFH has a direct impact on the inferred SFR and $M_\star$ \citep[e.g.,][]{carnall2019a}. Two main approaches are commonly used to describe galaxy SFHs: nonparametric and parametric \citep[e.g.,][for a comparative approach]{carnall2019a, leja2019a}. Nonparametric SFHs are constructed from statistical distributions that describe the variation of the SFR across successive age bins, whereas parametric SFHs rely on analytic functions characterized by a few parameters. Given the size of our sample and the much greater computational cost of nonparametric modeling, we adopted a parametric SFH.

Popular parametric forms include log-normal or double power-law functions. However, such commonly used models have been shown to inadequately reproduce the SFMS, as they fail to account for rapidly quenched or star-bursting systems, and introduce artificial age gradients along the main sequence not seen in simulations \citep{ciesla2017a}. To address these limitations, we adopted a more flexible formulation: a ``delayed" SFH ($\mathrm{SFR}\propto t\times e^{-t/\tau}$) modified by a recent burst or quenching episode modeled by a constant SFR. This configuration captures both the build-up of the bulk of the stellar mass via the delayed component and recent SFR variations, either upward or downward, traced by the blue and ultraviolet emission.

In CIGALE, this SFH is implemented through the \texttt{sfhdelayedbq}) module, which is governed by four parameters:
\begin{itemize}
    \item \texttt{age\_main}: age of the oldest star,
    \item \texttt{tau\_main}: $e$-folding timescale of the exponential component,
    \item \texttt{age\_bq}: look-back time to the onset of the most recent burst or quenching event,
    \item \texttt{r\_sfr}: ratio of the current SFR to that immediately preceding the burst or quenching event.
\end{itemize}

\subsubsection{Stellar populations}

As described in Sect.~\ref{ssec:SSP}, we employed the BC03 (module \texttt{bc03}) and CB19 (module \texttt{cb19}) SSP models, as well as BPASS (module \texttt{bpassv2)}, which accounts for both single and binary stellar evolution. For all models, we adopted a common \cite{salpeter1955a} IMF and considered two metallicities: subsolar ($Z=0.008$) and solar ($Z=0.02$).

\subsubsection{Nebular emission}

Young stellar populations efficiently ionize surrounding gas, producing nebular emission. We modeled this contribution with the \texttt{nebular} module, which includes both continuum processes (free-free, free-bound, and two-photon), potentially non-negligible for young populations at short wavelength, and emission lines from Hydrogen and metals, which can significantly affect broadband photometry in galaxies with strong star formation \citep[e.g.,][]{anders2003a}. As for the stellar populations, we adopted fixed metallicities of $Z=0.008$ and $Z = 0.02$.

\subsubsection{Dust attenuation}

Dust strongly influences galaxy emission through wavelength-dependent absorption and scattering. If not properly accounted for, dust attenuation results in underestimated $M_\star$ and, more critically, SFR, as it significantly suppresses UV emission from massive stars, the most direct tracer of recent star formation.

While the \cite{calzetti2000a} starburst attenuation curve is often adopted, numerous studies have demonstrated significant variations in attenuation curves among both nearby and distant galaxies \citep[e.g.,][]{salim2018a, salim2020a, boquien2022a}. One key factor driving these variations is the differential attenuation between stars of different ages. To account for this, we adopted the \cite{charlot2000a} two-component dust model (module \texttt{dustatt\_modified\_CF00}). In this framework, young massive stars are attenuated by dust within both their birth clouds and the diffuse interstellar medium (ISM), whereas older stellar populations experience attenuation solely from the ISM. Each component is assigned its own attenuation curve, where we used the slopes suggested by \citet{charlot2000a}-- $-0.7$ for the ISM and $-1.3$ for the birth clouds. This effect is particularly important for galaxies whose emission at short wavelengths is dominated by young stellar populations, as is expected for much of our sample.

This module has two free parameters:
\begin{itemize}
    \item \texttt{Av\_ISM}: the V-band attenuation in the ISM,
    \item \texttt{mu}: the ratio of the V-band attenuation in the ISM to the total V-band attenuation.
\end{itemize}
In effect, \texttt{mu} quantifies the dust geometry: values near 0 correspond to dust concentrated in star-forming regions, while values near 1 correspond to dust more uniformly distributed throughout the galaxy.

\subsection{Estimation of physical properties}
\label{ssec:estimation}

Using the modules described above, we constructed a Cartesian grid of models. To optimize computation time, the models were generated on a redshift grid with a step size of 0.01 rather than at the exact redshift of each galaxy. This approximation has a negligible impact on the fits, given the width of the broadband filters. The derived physical properties were, in any case, corrected to the exact input redshift. The full list of parameters used for each module is given in Table~\ref{tab:params}.

We adopted the same input grid for all galaxies in the sample, with one key exception: the age of the stellar populations (\texttt{age\_main}) was set to the age of the Universe at the redshift of each galaxy minus a few hundred Myr. This choice reflects that \texttt{age\_main} is difficult to constrain, particularly for strongly star-forming galaxies \citep[][and references therein]{Pforr2012}, where the light from young stellar populations dominates over that from older stars. As a result, unconstrained fits can yield unrealistic young ages compared to the cosmic age at the observed redshift. Assuming that the first stars formed a few hundred million years after the Big Bang naturally mitigates this issue and produces galaxy ages consistent with our current understanding of early galaxy formation \citep[e.g.,][]{iyer2017}. 

The metallicity was held fixed for each run to avoid degeneracies, but we performed the analysis for two representative metallicities: solar ($Z=0.02$) and subsolar ($Z=0.008$), motivated by the lower metal content typically observed in high-redshift galaxies \citep[e.g.,][]{faisst2016a}.

To account for unknown systematic uncertainties in both the observed fluxes and the models, we added an additional 5\% uncertainty in quadrature to all the fluxes. This approach, commonly adopted in SED modeling studies \citep[e.g.,][]{noll2009a, nersesian2019a}, balances the need for flexibility against the risk of overestimating errors, which could yield artificially low reduced $\chi^2$ values (hereafter noted as $\chi_r^2$). Our choice is also consistent with recent JWST-based analyses that impose a 5\% uncertainty floor \citep[e.g.][]{alberts2024a, helton2025a}.

We also treated measurements with signal-to-noise ratios (S/N) < 2 as non-detections and considered them as upper limits. A detailed discussion on how CIGALE handles upper limits is given in \citet{boquien2019a}.  

\begin{figure}[!ht]
    \centering
    \includegraphics[width=\linewidth]{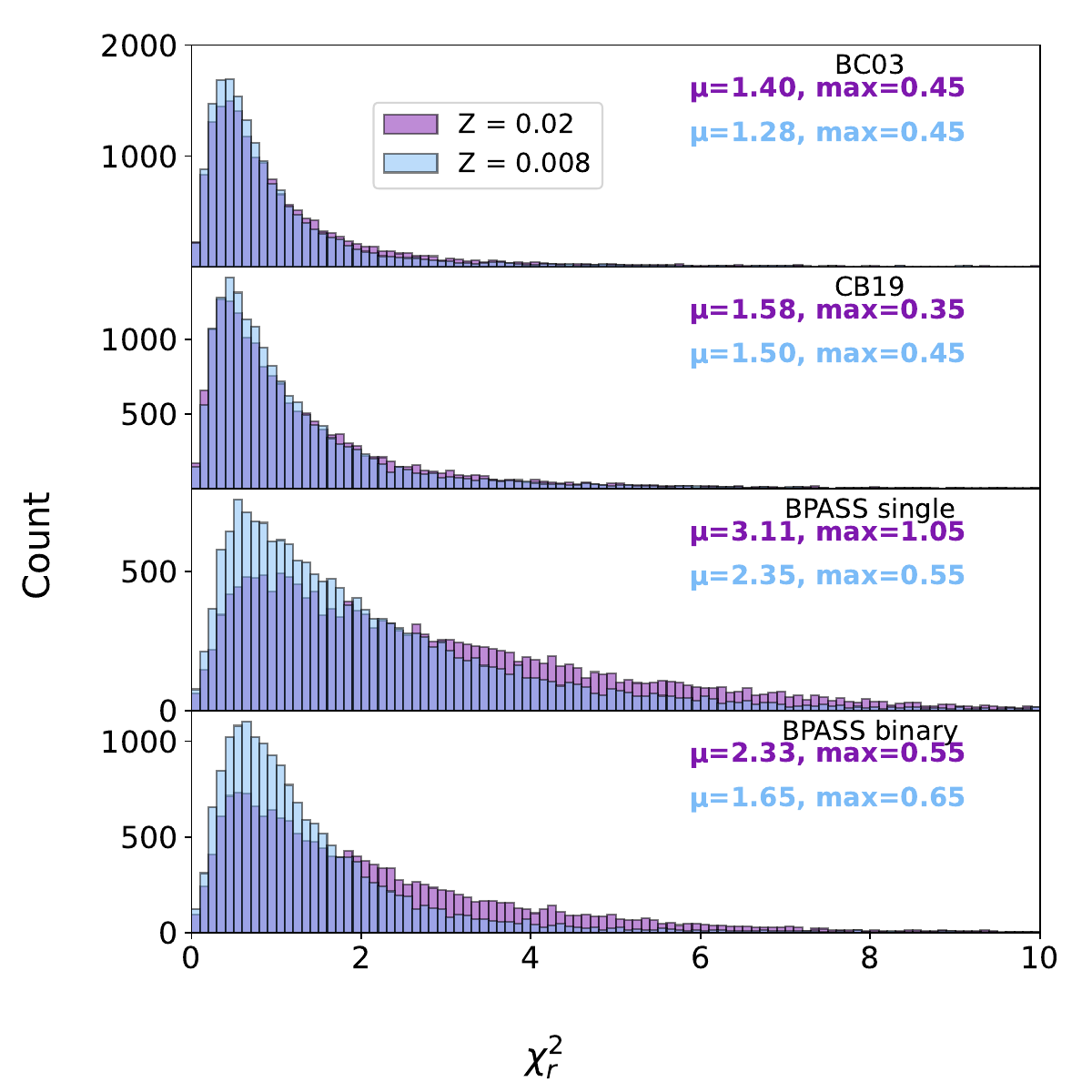}
    \caption{Distribution of the $\chi_r^2$ values of the best-fit models. The distribution peaks around $0.5$ with a long tail toward higher values and a mean value generally between 1 and 3. In this case, the histograms are limited to $0.0<\chi_r^2<10.0$ for clarity. A small number of objects have higher values.
    }
    \label{fig:chi-obs}
\end{figure}

For each object, CIGALE fits the model grid to the observed photometry, estimating the physical properties and their uncertainties from the likelihood-weighted means and standard deviations. Fig.~\ref{fig:chi-obs} shows the distribution of the $\chi_r^2$ values of the best-fit model for each of the four stellar population models considered.

Overall, the fits are satisfactory, with mean $\chi_r^2$ values typically between 1 and 3, depending on the model and metallicity. The distribution peaks slightly below unity, which likely reflects variations in the effective number of degrees of freedom, which is difficult to estimate precisely and should ideally be determined experimentally for each object \citep[e.g.,][]{smith2012a}. Given the large sample size and the limited impact on our conclusions, we retained this approximate treatment.

\begin{table}
\caption{Median uncertainties in the inferred SFR and $M_\star$ for the sample of observed galaxies.}
\label{tab:errors}
\centering
\begin{tabular}{l|l|l|l}
$Z$                    & Model        & $\sigma_{\mathrm{SFR}}$ {[}dex{]} & $\sigma_{M_\star}$ {[}dex{]} \\ \hline
\multirow{4}{*}{0.02}  & BC03         & 0.28                                   & 0.04                                  \\
                       & CB19         & 0.28                                   & 0.04                                  \\
                       & BPASS single & 0.12                                   & 0.05                                  \\
                       & BPASS binary & 0.19                                   & 0.06                                  \\ \hline
\multirow{4}{*}{0.008} & BC03         & 0.23                                   & 0.06                                  \\
                       & CB19         & 0.30                                   & 0.06                                  \\
                       & BPASS single & 0.15                                   & 0.06                               \\                     
                       & BPASS binary   & 0.21                                 & 0.06
                       
\end{tabular}
\end{table}

\begin{figure*}[!ht]
    \centering
    \includegraphics[width=\columnwidth]{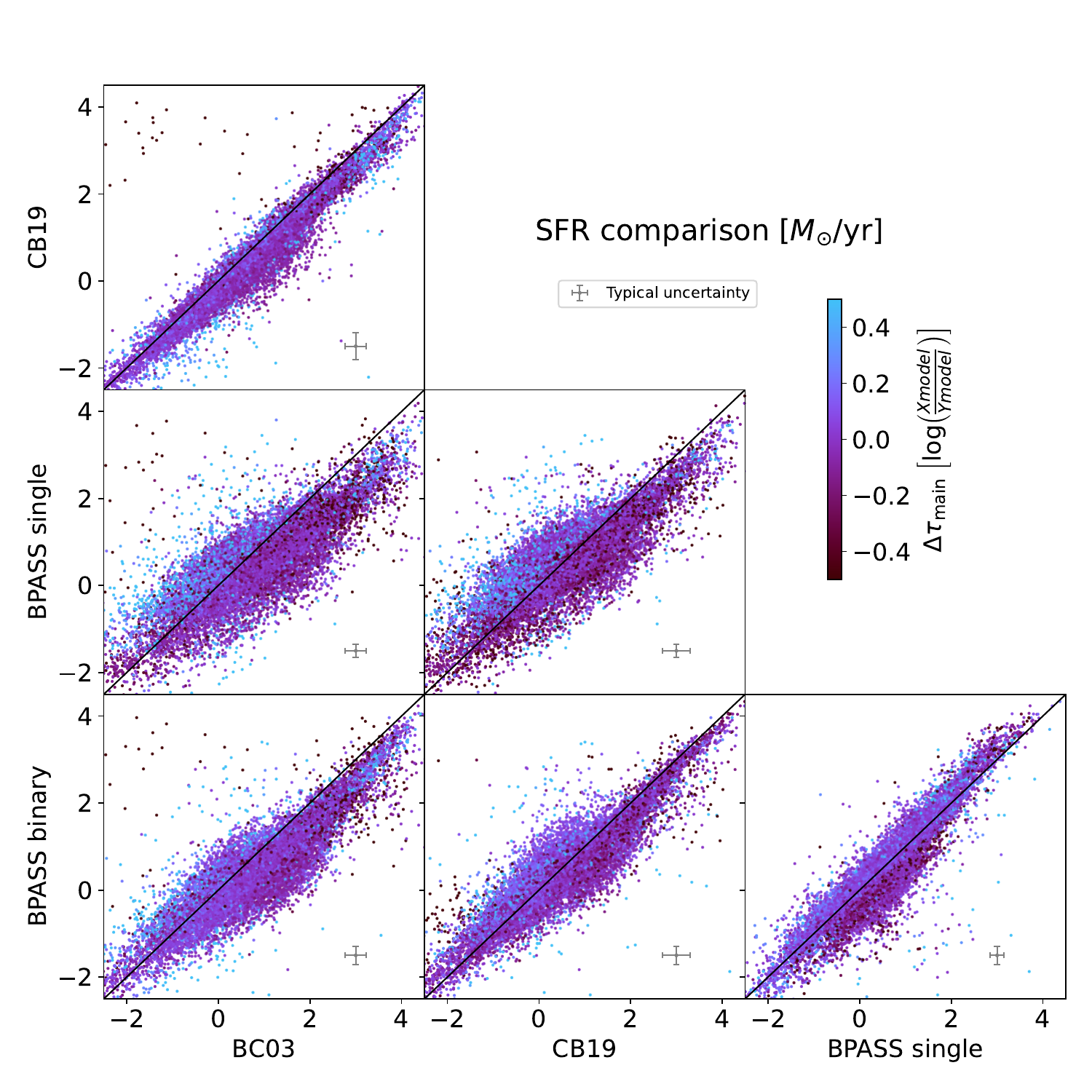}
    \includegraphics[width=\columnwidth]{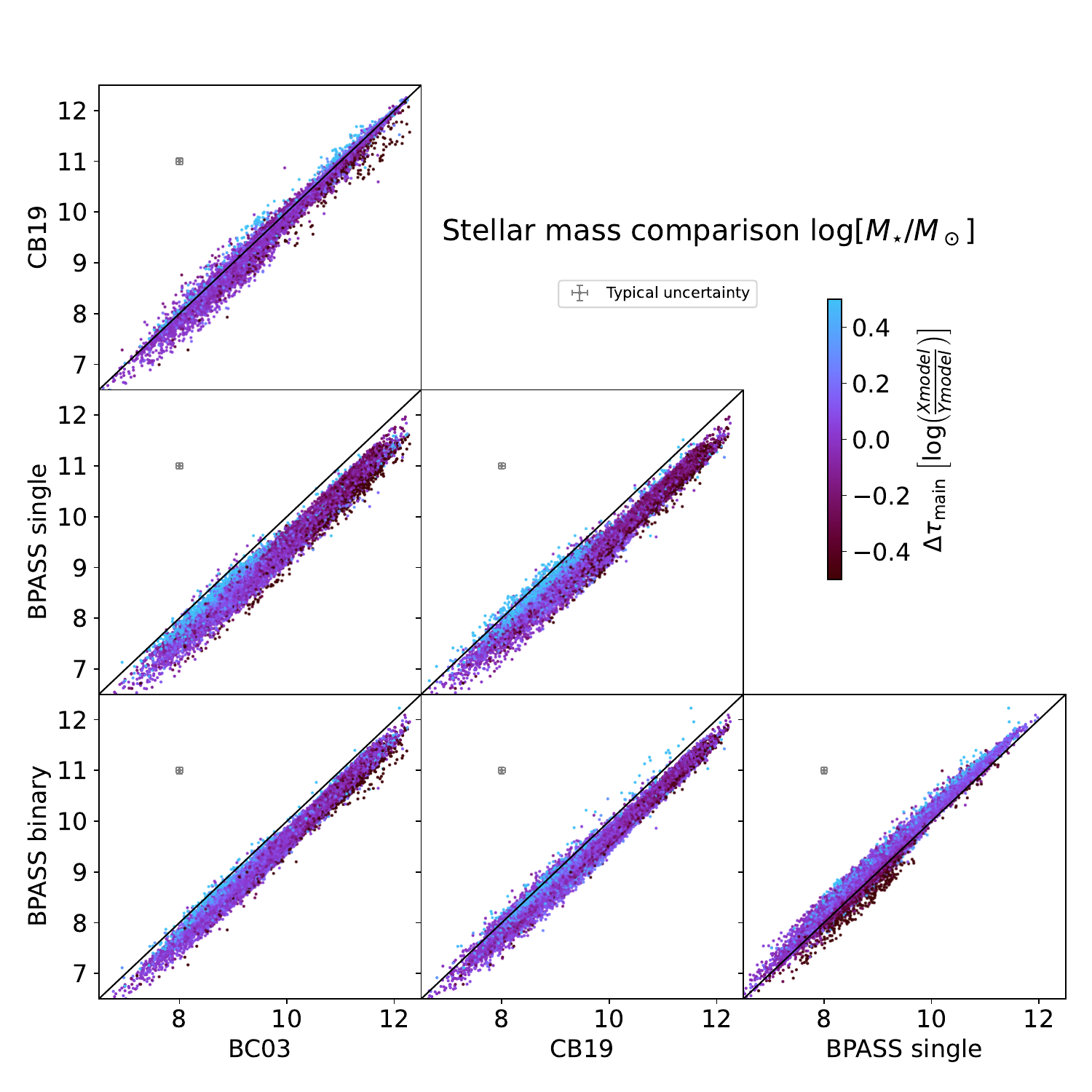}
    \caption{Plot of the comparison of the SFR (left) and $M_\star$ (right) using four different stellar population models with CIGALE. The median relative uncertainty is shown in the top left of every plot. The solid black lines correspond to $x=y$. These results were obtained using the parameters shown in Table~\ref{tab:params}. These plots show the results for $Z=0.008$. Similar plots for $Z=0.02$ are shown in Figure~\ref{fig:obs_02}.
    }
    \label{fig:obs}
\end{figure*}

The typical uncertainties on SFR and $M_\star$ for individual objects are $\sim 0.2$~dex and $\sim 0.05$~dex, respectively (see Table~\ref{tab:errors}). Fig.~\ref{fig:obs} shows pairwise comparisons of SFR and $M_\star$ obtained with the different models, revealing overall qualitative agreement but with systematic offsets that we discuss in the following sections. These plots are also colored by the difference in one of the derived SFH parameters, namely $\tau_\text{main}$. This is to showcase the differences in the inferred SFH induced by the change of the SPS model. While larger $\tau_{\text{main}}$ values naturally correspond to higher SFRs, we find that differences between model families in $\Delta \tau_{\text{main}}$ can correspond to reduced offsets in $M_\star$.

\subsection{Construction and modeling of synthetic catalogs}
\label{sec:synth_phot}

Modeling the observed sample, as described in the previous section, informs us about the relative differences in the derived physical properties when using different stellar population models. However, this approach provides limited insight into possible intrinsic biases introduced by the models themselves. To address this question, we performed a second analysis based on synthetic photometric catalogs.

Specifically, we generated four sets of synthetic catalogs, each corresponding to one of the stellar population models used in this study. For each galaxy, we used its best-fit model from the observed sample as the ``ground truth'' and computed the corresponding noiseless fluxes in all bands. By construction, these synthetic galaxies have perfectly known physical properties and fluxes, which are otherwise inaccessible for the observed sample. The resulting synthetic catalogs preserve the overall statistical properties of the observed galaxies, ensuring it is representative of the parent population.

To ensure that the synthetic photometry remains realistic, we added noise to the best-fit fluxes of each galaxy but not to the upper limits, which were left as is. This noise was drawn from a Gaussian distribution with a standard deviation equal to the total uncertainty, including the added 5\% uncertainty discussed in Section~\ref{ssec:estimation}. The same uncertainties were then adopted as the flux uncertainties in the synthetic catalogs. 

\begin{figure*}[!htbp]
    \centering
    \includegraphics[width=\columnwidth]{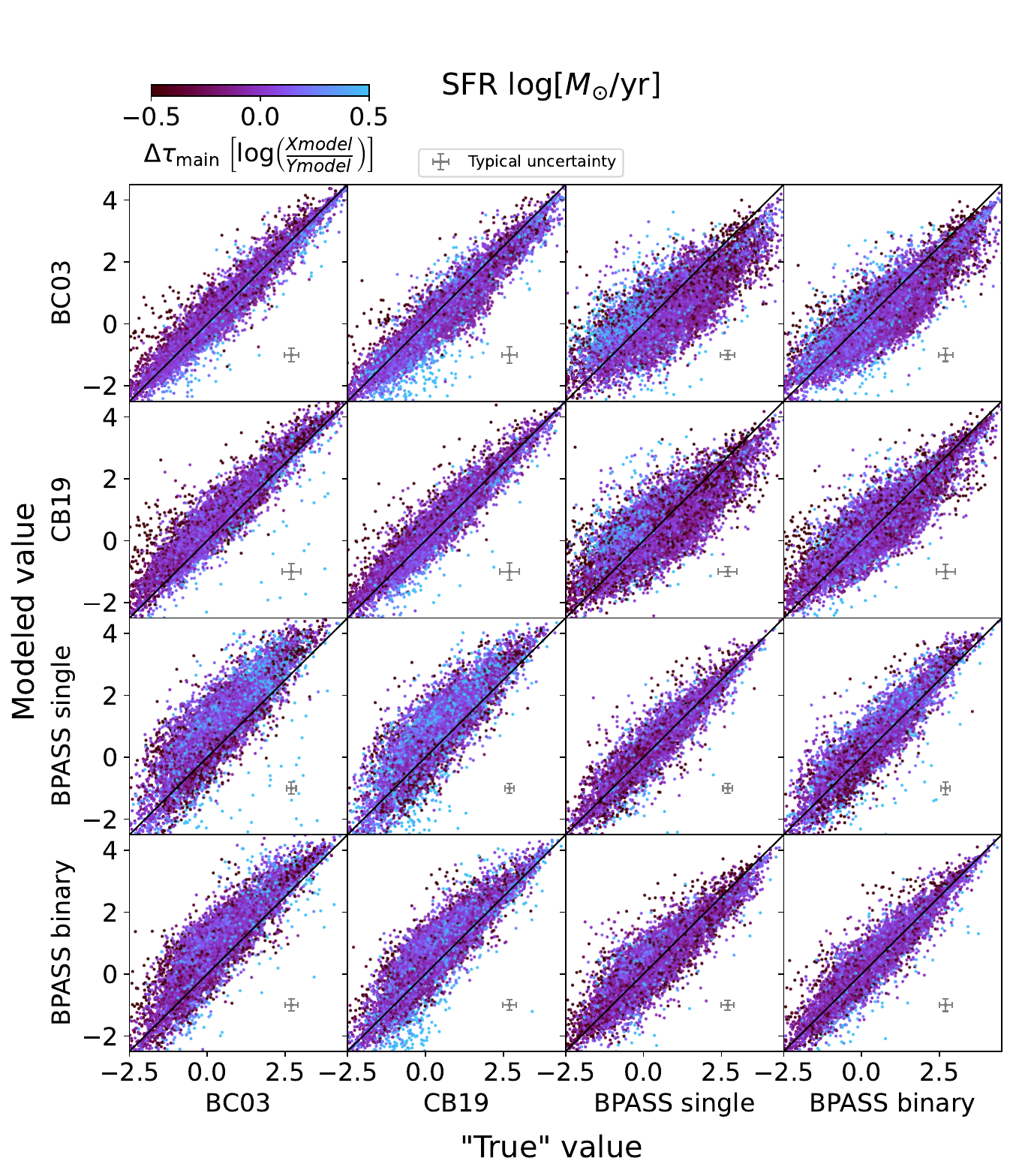}
    \includegraphics[width=\columnwidth]{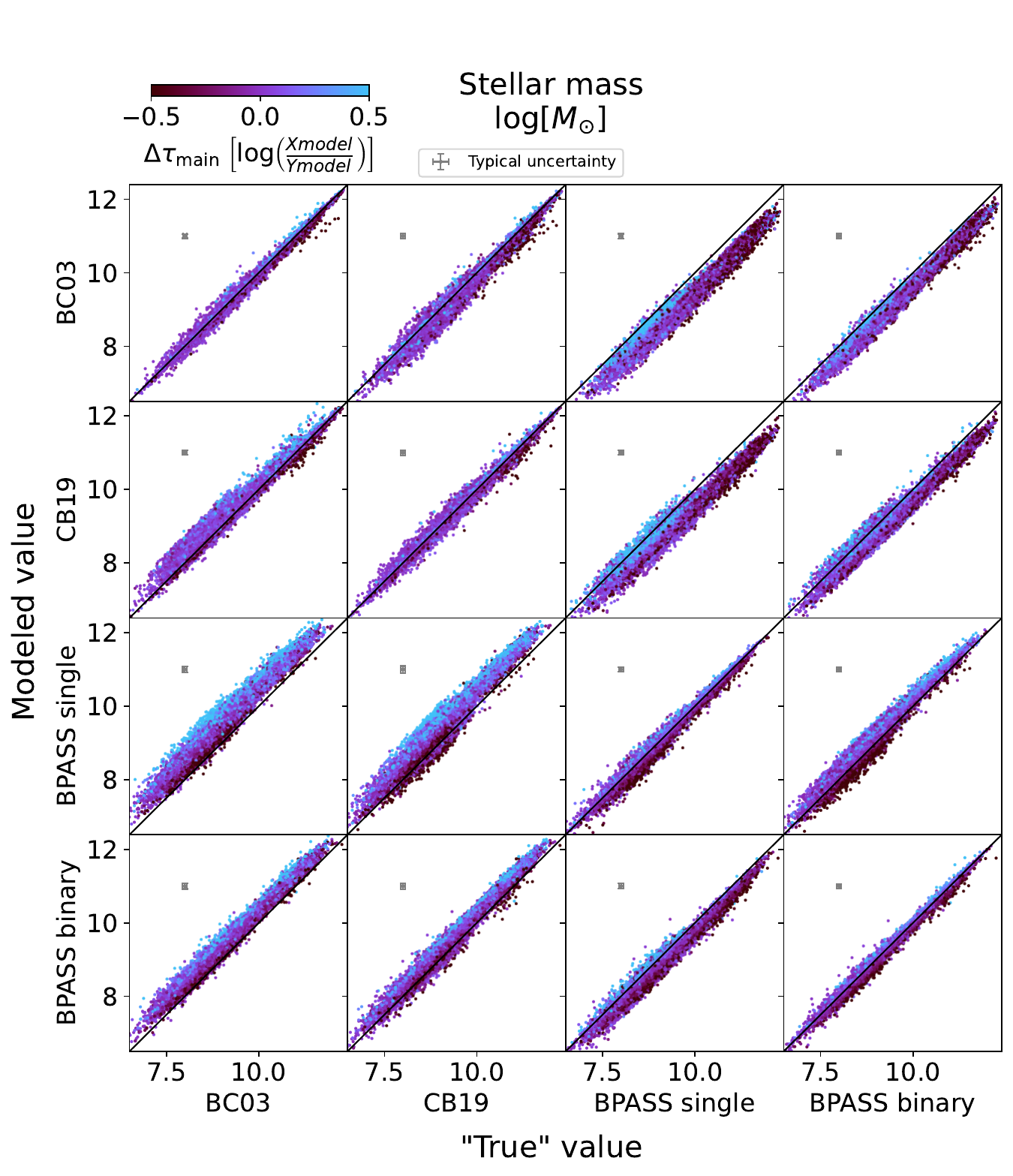}
    \caption{Comparison of the ``true'' ($x$-axis) to the estimated ($y$-axis) SFR (left) and $M_\star$ (right) for the galaxies in the synthetic catalog for every stellar population. The median relative uncertainty of the sample is shown in the top left of each plot. The solid black lines correspond to a one-to-one relation. These results were obtained using the parameters shown in Table~\ref{tab:params}. These plots show the results for $Z=0.008$. Similar plots for $Z=0.02$ are shown in Figure~\ref{fig:synth_02}.}
    \label{fig:synth}
\end{figure*}

Each synthetic catalog was then fitted with all four stellar population models in turn (e.g., galaxies generated with BC03 populations were also fitted with CB19 models), using the same model grids as for the observed sample. This cross-analysis allows us to quantify systematic biases that arise solely from the intrinsic differences between the stellar population models, disentangled from observational uncertainties or calibration systematics. We compare the estimated and true SFR and $M_\star$ in Fig.~\ref{fig:synth}. As for the observed galaxies, we find there is an overall qualitative agreement between models but also clear systematic deviations, which we discuss in detail below.

\section{Results\label{sec:results}}

\subsection{Comparison of inferred physical properties from four stellar population models}

In Fig.~\ref{fig:obs}, we compare estimates of SFR and $M_\star$ obtained using four SSP models for a sample of 17,230 galaxies. The estimates are strongly correlated, indicating overall consistency among the models. A qualitative inspection reveals no major systematic bias between models within the same family, that is, between BC03 and CB19, and between BPASS single and BPASS binary. This suggests that the main source of systematic variation arises from differences between those two families, rather than from binary evolution alone.

\begin{figure*}[!htbp]
    \includegraphics[width=\columnwidth]{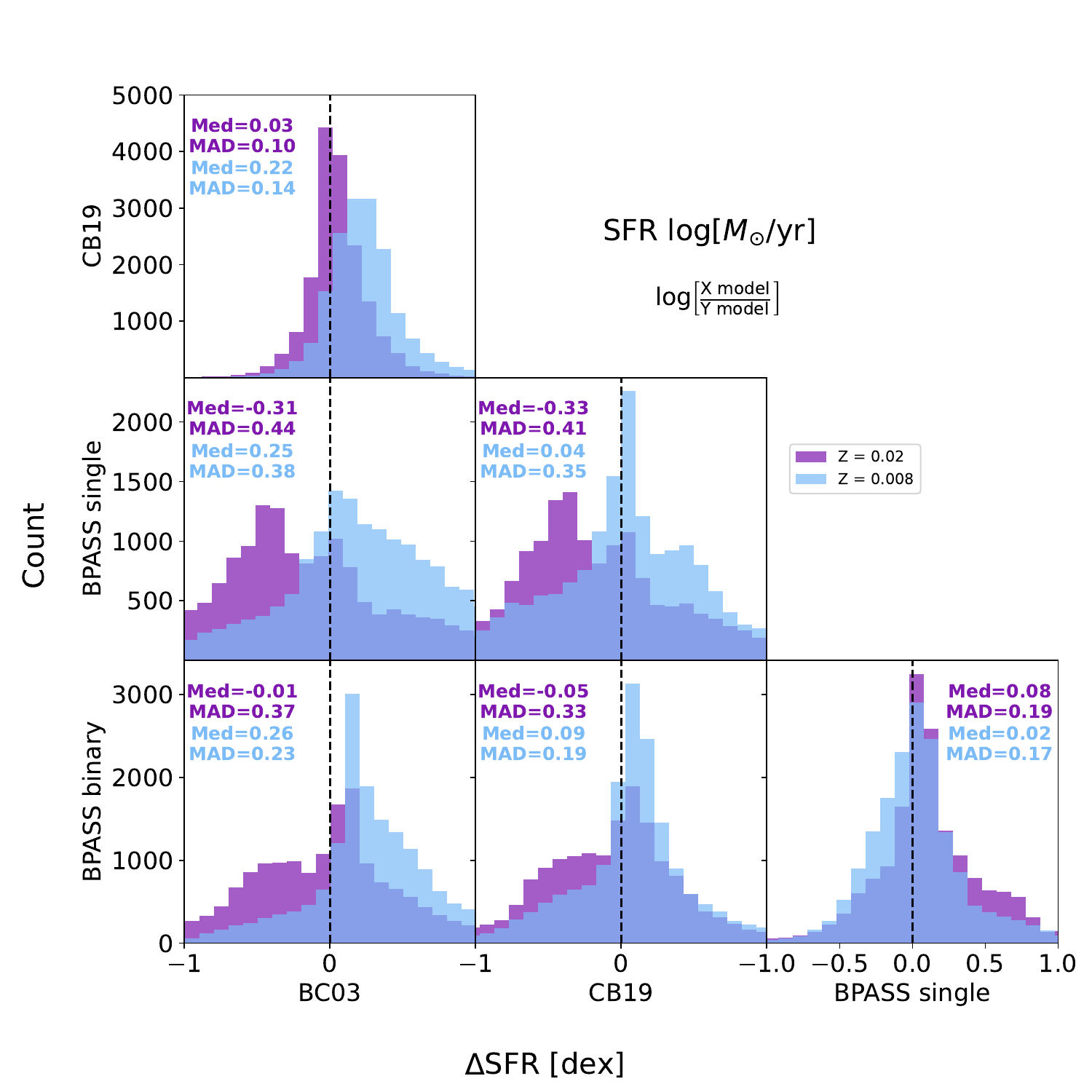}
    \includegraphics[width=\columnwidth]{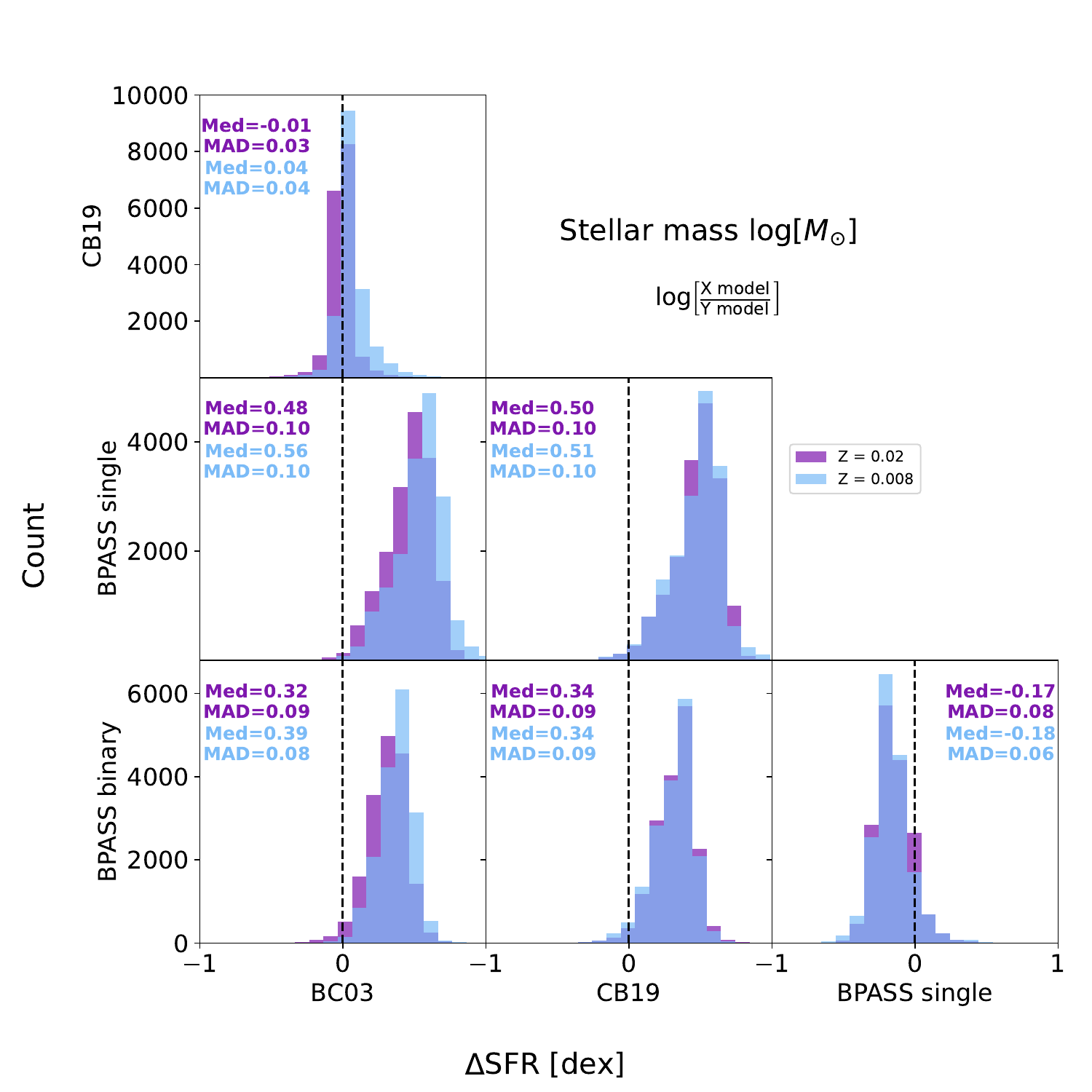}
    \caption{Histograms of the offsets in dex between the physical properties (left: SFR, right: $M_\star$) determined by the model on the $y$-axis label and the model on the $x$-axis label inferred from the fits on the observed fluxes. The order of the plots follows Fig.~\ref{fig:obs}. The dashed line shows the location of 0.0~dex offset. The blue histograms are for $Z=0008$ and the purple for $Z=0.02$. Every plot shows the respective median and median absolute deviation of the offset.
    }
    \label{fig:offset-obs}
\end{figure*}

To quantify the offsets, we show in Fig.~\ref{fig:offset-obs} the histograms of the offsets between models.

Some offset is present in most cases, but the BC03 and CB19 models are the most consistent with each other, while both differ more significantly from the BPASS models. In general, the widely accepted BC03 models yield the highest masses and SFRs among the four.

It is informative to compare the relative amplitudes of these offsets with the median fitting uncertainties (Table~\ref{tab:errors}). The offsets between the different model families for $M_\star$ are typically larger than the median uncertainties, indicating that systematic model differences dominate. Even within a single family, the offset between the BPASS single and binary models exceeds the median uncertainty on $M_\star$. By contrast, the BC03--CB19 offset is comparable to the fitting uncertainty, implying that systematic and random effects are of similar importance in that case. For SFR, the offsets between models are generally smaller than or comparable to the typical uncertainty, indicating that SFR estimates are less sensitive to the stellar population models. These conclusions hold for both tested metallicities.

\subsection{Model inter-comparison through synthetic catalogs}

To further assess the relative impact of the choice of the SSP model, we constructed four synthetic catalogs, each based on the best fits obtained with one of the four models. We then estimated the physical properties of each synthetic catalog using all four SSP models.

This approach offers two key advantages. First, fitting a synthetic catalog with the same SSP model used to generate it naturally removes uncertainties due to the SSP itself, isolating the effects of other modeling assumptions (e.g., the adopted SFH and attenuation laws) and of the photometric configuration (e.g., wavelength coverage, depth). This represents the best accuracy achievable for a given setup. Second, when fitting the synthetic photometry generated with one model using a different model, the differences in the SSP induce differences in the derived physical properties, such as the SFH. Such differences are a good indicator of the uncertainties on physical properties stemming from the SSPs themselves, thus allowing us to quantify systematic and random biases attributable to the SSP models.

\begin{figure*}[!htbp]
    \includegraphics[width=\columnwidth]{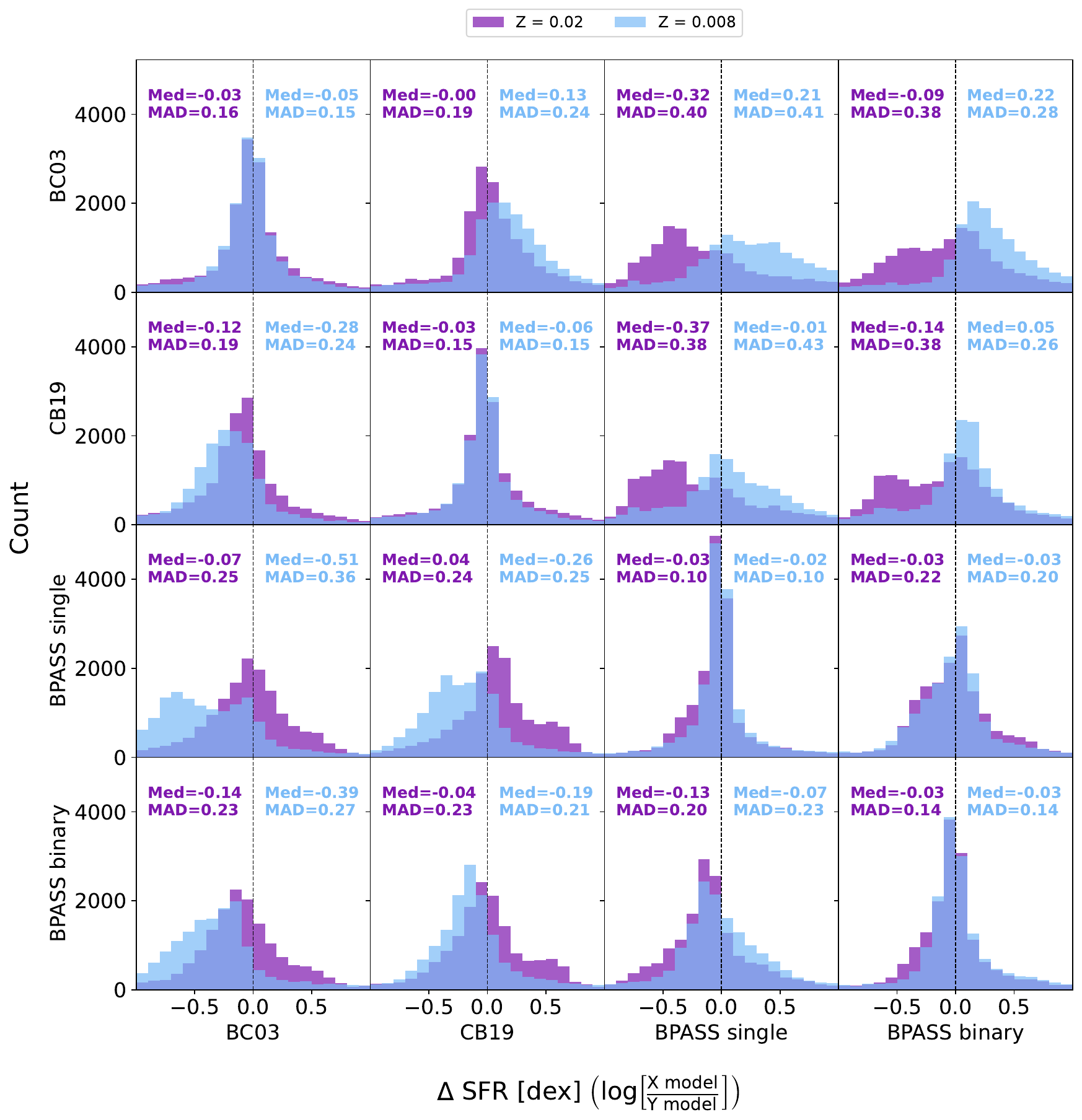}
    \includegraphics[width=\columnwidth]{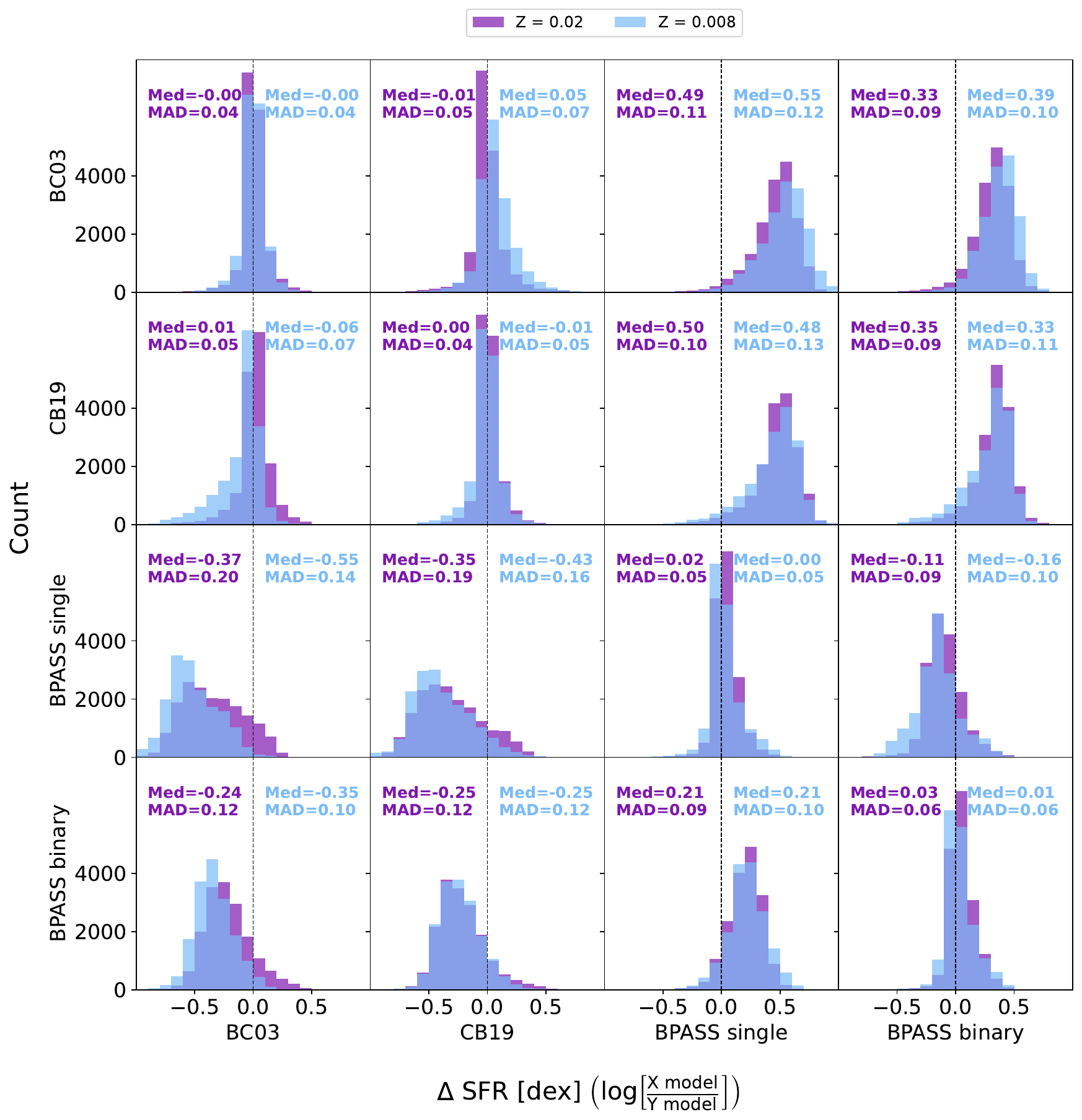}
    \caption{Histograms showing the offsets between the estimated and the ``true'' values for SFR (left) and $M_\star$ (right). Every column shows the results on a different set of synthetic data (generated as described in Fig.~\ref{sec:synth_phot}).  The blue histograms are for $Z=0.008$ and the purple for $Z=0.02$. The CIGALE parameters used for every run are the same as for the observed data and as shown in Table.~\ref{tab:params}.}
    \label{fig:offset-synth}
\end{figure*}

Fig.~\ref{fig:synth} shows the comparison between the ``true'' SFR and $M_\star$ and the estimates obtained from the fits. To facilitate comparison, Fig.~\ref{fig:offset-synth} presents the corresponding histograms of offsets.
For each histogram, the median and median absolute deviation (MAD) are shown. The diagonal panels, corresponding to each model fitted to its own synthetic catalog, show small median offsets, no more than $0.06$~dex for SFR and $0.03$~dex for $M_\star$. The MAD is systematically larger for SFR (up to $0.16$~dex than for $M_\star$ (up to $0.06$~dex)).

The off-diagonal panels, where one model is fitted to the synthetic photometry generated by another, show median offsets and scatters that are overall comparable to those observed in the real data. For example, the $M_\star$ offset between the BC03 and the BPASS binary models ($Z = 0.008$) is $0.32 (0.09)$~dex for the observed fluxes (Fig.~\ref{fig:offset-obs}) and $-0.35 (0.10)$~dex and $0.39 (0.10)$~dex for the respective synthetic catalogs (Fig.~\ref{fig:offset-synth}). This indicates that, given the photometric coverage and depth of our dataset, the dominant source of the offsets lies in the intrinsic differences between the SSP models.   

To quantify the amplitude of these systematic offsets, Tables~\ref{tab:offsets-sfr} and \ref{tab:offsets-mass} list the median offsets for SFR and $M_\star$, respectively. For each model, the offset derived from the self-fit case (same model for generation and fitting) was subtracted to isolate the contribution of the SSP model differences.

\begin{table*}[!ht]
\caption{SFR offset in dex that is due to differences in the stellar population models.}
\label{tab:offsets-sfr}
\centering
\begin{tabular}{c|c|c|c|c|c}
\backslashbox{Synthetic catalog model}{Fitting model}   & BC03    & CB19    & BPASS single & BPASS binary   & $Z$                   \\ \hline
BC03                                                    & $+0.00$ & $-0.09$ & $-0.09$      & $-0.11$        &\multirow{4}{*}{0.02}  \\
CB19                                                    & $+0.03$ & $+0.00$ & $+0.07$      & $-0.01$        &                       \\
BPASS single                                            & $-0.29$ & $+0.34$ & $+0.00$      & $-0.10$        &                       \\
BPASS binary                                            & $-0.06$ & $-0.11$ & $+0.00$      & $+0.00$        &                       \\ \hline
BC03                                                    & $+0.00$ & $-0.23$ & $-0.46$      & $-0.34$        &\multirow{4}{*}{0.008} \\
CB19                                                    & $+0.19$ & $+0.00$ & $-0.20$      & $-0.13$        &                       \\
BPASS single                                            & $+0.23$ & $+0.01$ & $+0.00$      & $-0.05$        &                       \\
BPASS binary                                            & $+0.25$ & $+0.08$ & $+0.00$      & $+0.00$        &                      

\end{tabular}
\tablefoot{Offsets originating from the modeling and photometric coverage have been subtracted, using the median values shown in Figure~\ref{fig:offset-synth}. Each combination of fitting model and synthetic catalog model - the offset when the fitting model and the synthetic catalog model coincide.}
\end{table*}

\begin{table*}[!ht]
\caption{Same as in Table~\ref{tab:offsets-sfr} for $M_\star$.}
\label{tab:offsets-mass}
\centering
\begin{tabular}{c|c|c|c|c|c}
\backslashbox{Synthetic catalog model}{Fitting model} & BC03 & CB19 & BPASS single & BPASS binary & $Z$                   \\ \hline
                                        BC03         & $+0.00$ & $+0.01$ & $-0.37$ & $-0.24$ &\multirow{4}{*}{0.02}  \\
                                        CB19         & $-0.01$ & $+0.00$ & $-0.35$ & $-0.25$ &\\
                                        BPASS single & $+0.47$ & $+0.48$ & $+0.00$ & $+0.19$ &\\
                                        BPASS binary & $+0.30$ & $+0.32$ & $-0.14$ & $+0.00$ &\\ \hline
                                        BC03         & $+0.00$ & $-0.06$ & $-0.55$ & $-0.35$ &\multirow{4}{*}{0.008} \\
                                        CB19         & $+0.05$ & $+0.00$ & $-0.43$ & $-0.25$ &\\
                                        BPASS single & $+0.55$ & $+0.48$ & $+0.00$ & $+0.21$ &\\
                                        BPASS binary & $+0.38$ & $+0.32$ & $-0.17$ & $+0.00$ &
\end{tabular}
\end{table*}

For SFR, the resulting offsets are smaller than the internal dispersion due to photometric and modeling uncertainties, confirming that the choice of a SSP model has a limited impact on SFR estimates. 

In contrast, the offsets in $M_\star$ are significantly larger, reaching $0.55$~dex between the BC03 model and BPASS single. The offsets between model families are generally larger than or comparable to the intrinsic dispersion, while those within a family remain smaller, up to $0.06$~dex between BC03 and CB19, and up to $0.21$~dex between BPASS single and BPASS binary.

These findings are broadly consistent with previous studies. Using CEERS data \citep{finkelstein2025a}, \cite{osborne2024a} find that BC03 yields higher $M_\star$ values than BPASS by $\sim0.17$~dex, slightly below our estimate ($0.24[0.30] (Z = 0.02)$~dex and $0.35 [0.38](Z = 0.008)$~dex for BC03 synthetic data [model] and BPASS binary model [synthetic data]). However, the internal dispersions are large (MAD$\sim0.1$) in all cases. Similarly, \cite{seille2024a}, analyzing $1<z<3$ star-forming galaxies from CEERS,  report results consistent with \cite{osborne2024a}.

\begin{figure}[!ht]
    \centering
    \includegraphics[width=\linewidth]{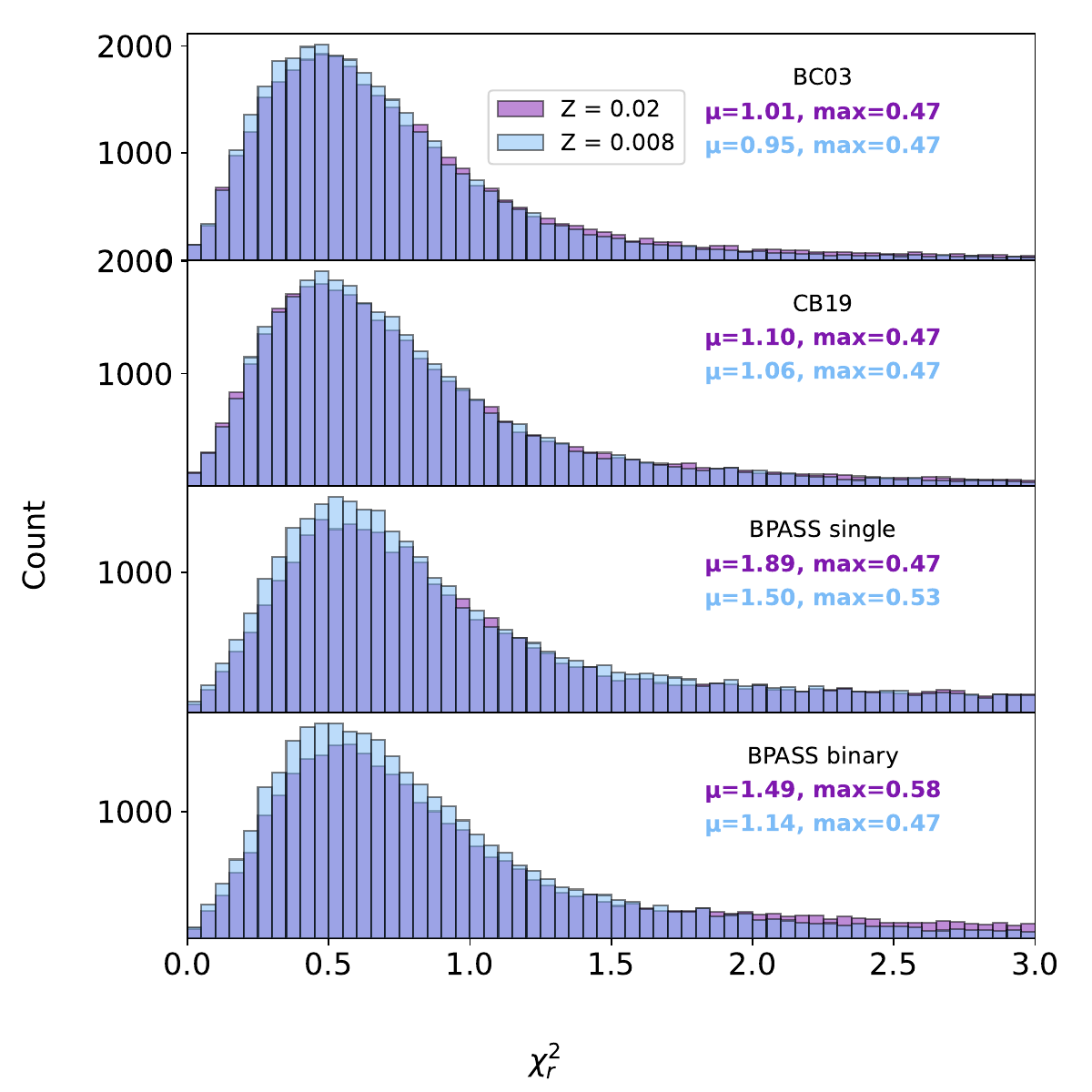}
    \caption{Distribution of $\chi_r^2$ on the synthetic catalogs when using the same stellar population models for the generation of the catalog and the fitting to estimate the physical properties.}
    \label{fig:chi2_synth}
\end{figure}

Overall, despite similar $\chi_r^2$ distributions (Fig.~\ref{fig:chi2_synth}), the differences in SFR and especially in $M_\star$ are non-negligible.
This shows that all the stellar population models are sufficiently flexible to reproduce each other's photometry, given the diversity of SFHs and attenuation curves explored, yet they yield systematically different physical parameters.

\section{Discussion\label{sec:discussion}}

\subsection{Predicted $M_\star/L_i$ ratios}

While a full treatment of the origin of the offsets quantified in Section~\ref{sec:results} is beyond the scope of this work, the dominant effects can be understood through a simplified argument relating observables to derived physical properties for the different SPS libraries. We focused on the stellar mass, as the systematic offsets do not dominate over other uncertainties when inferring the SFR. We addressed this by plotting the $g-i$ color against the $M_\star/L_i$ ratio predicted by the different SSP models \citep[see e.g.,][]{taylor2011}. The SEDs used here were generated with CIGALE, using only an SFH and a stellar population module (i.e., no dust or nebular emission). We adopted a double exponential SFH with a very short timescale $(\tau_{\text{main}} = 1)$~Myr, effectively tracing the evolution of a single burst over a range of ages.

\begin{figure}
    \centering
    \includegraphics[width=\linewidth]{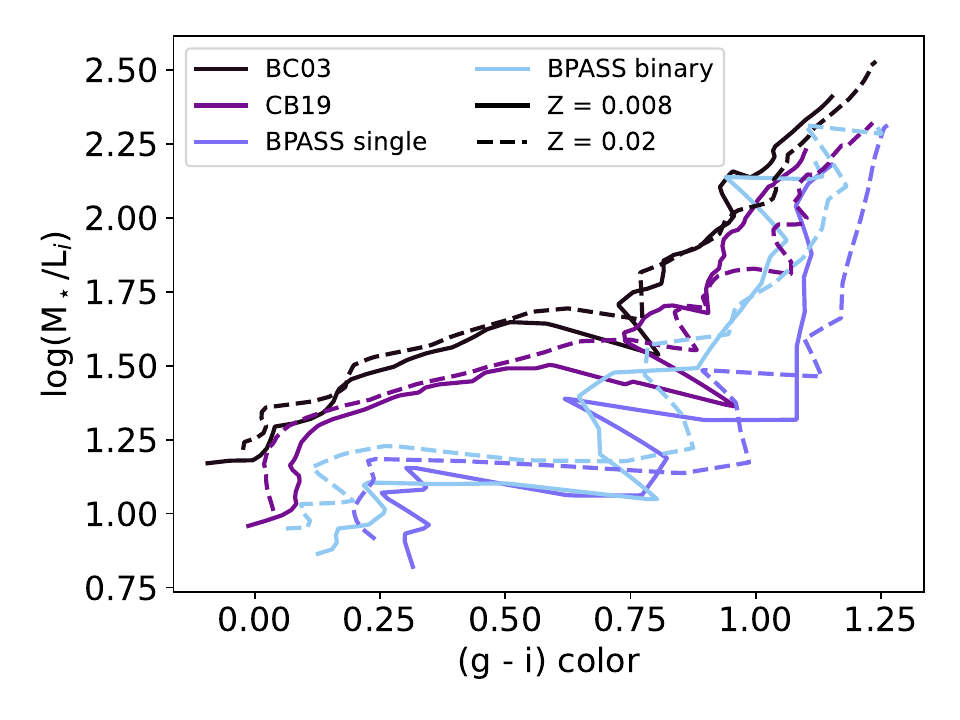}
    \caption{$M_\star/L_i$ as a function of color for the four SPS models. Solid and dashed lines correspond to $Z = 0.02$ and $ Z = 0.008$, respectively.}
    \label{fig:giML}
\end{figure}

The resulting relation (Fig.~\ref{fig:giML}) shows that the different SPS models predict systematically different mass-to-light ratios, with BC03 yielding the highest $M_\star/L_i$ at a fixed color. This implies that a given $g-i$ occurs at an older age in BC03. This offset is primarily driven by differences in the underlying single-star stellar evolution prescriptions, while the small separation between the single and binary BPASS models indicates that binary evolution plays a secondary role.

\subsection{Redshift dependence of the offsets}

\begin{figure*}[!ht]
    \centering
    \includegraphics[width=0.48\linewidth]{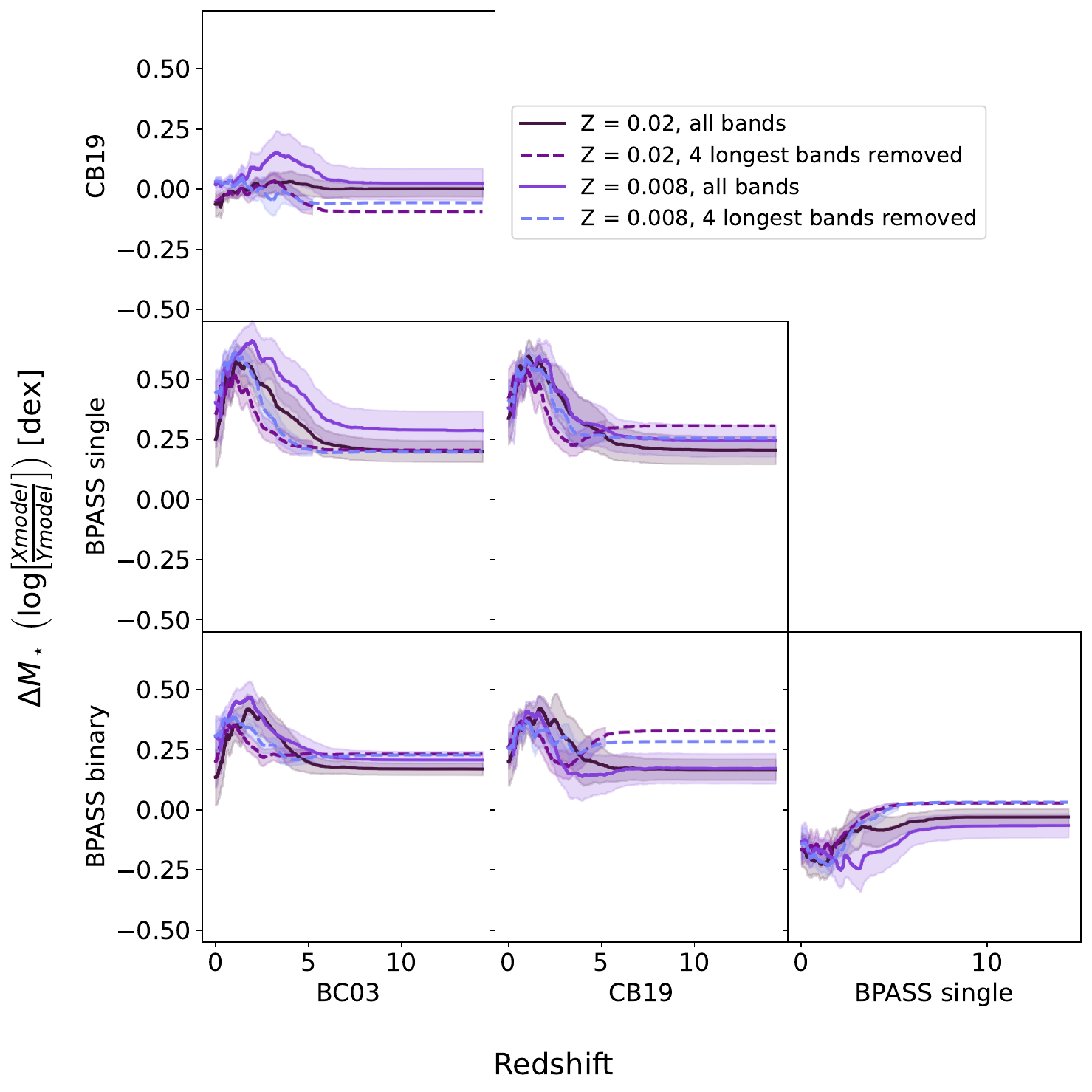}
    \includegraphics[width=0.48\linewidth]{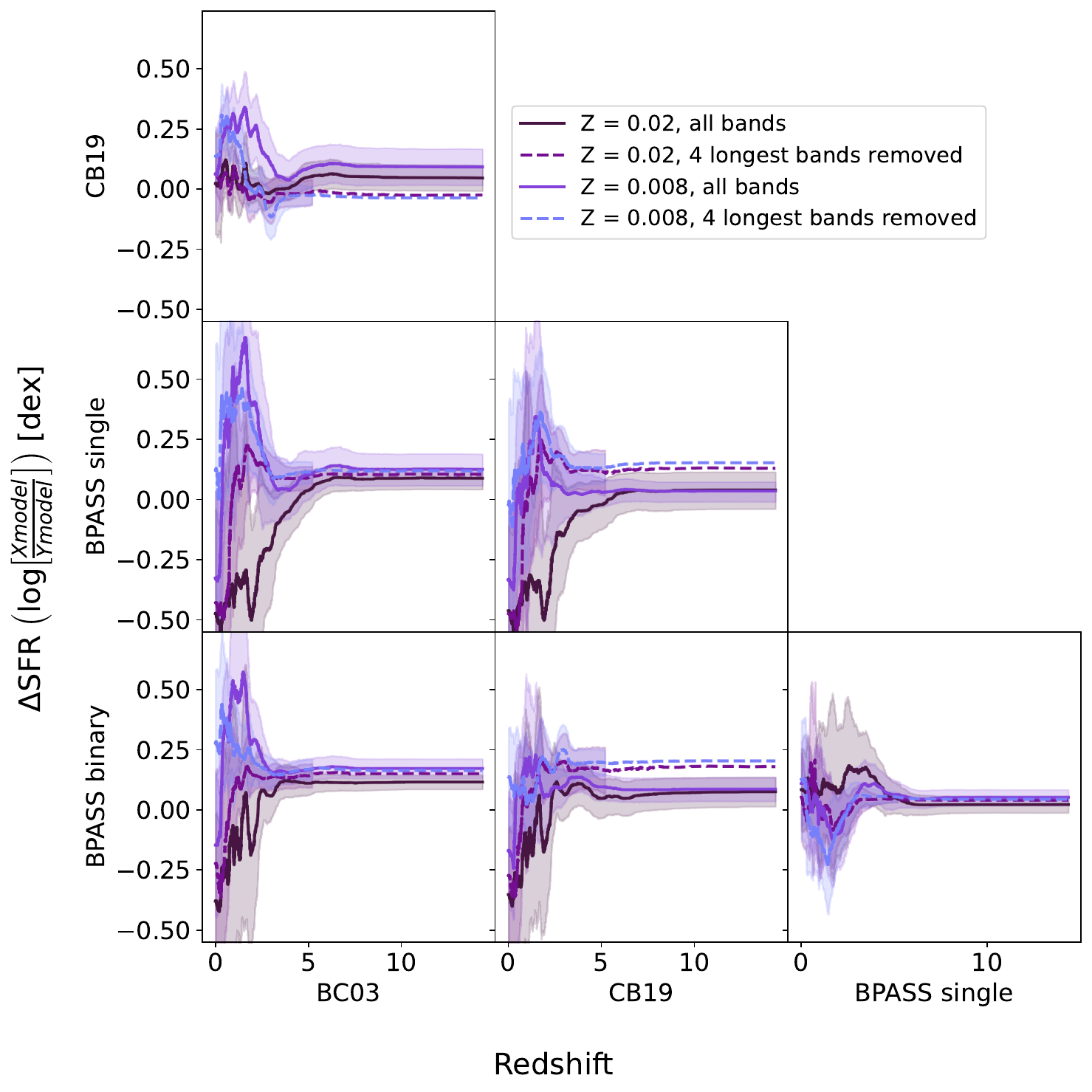}
    \caption{Moving average of the offsets between models in the inferred physical properties of the sample. The moving window was 1000 galaxies wide. The shaded regions show the MAD of the distribution in each window. Left: $M_\star$. Right: SFR.
    }
    \label{fig:mvav}
\end{figure*}

The redshift dependence of the offsets was examined by computing the moving average of $M_\star$ differences as a function of redshift, using a sliding window containing 1000 sources (Figure~\ref{fig:mvav}) and by plotting 2D histograms (Figure~\ref{fig:2dhist_mass} and Figure~\ref{fig:2dhist_sfr}) of the offsets with the redshift.

Within each model family, the offsets of the inferred stellar mass can reach values comparable to the average values shown in Table~\ref{tab:offsets-mass}. The offsets between model families follow a similar trend but can get larger at their peak (up to $\sim 0.6$~dex for BC03 and BPASS single-star). The $M_\star$ offsets in all cases gradually rise to a peak around $z\sim2$ after which they decrease and reach a near constant value after $z\sim5$. The size of the window used for calculating the moving average in the left panel of Figure~\ref{fig:mvav} was 1000 galaxies, but the same general trends are also seen when we used a smaller window (e.g., 100 galaxies). Both metallicities follow a similar trend. The trends for the SFR offsets are also similar, although generally noisier. 

\begin{figure}[!htbp]
    \centering
    \includegraphics[width=\linewidth]{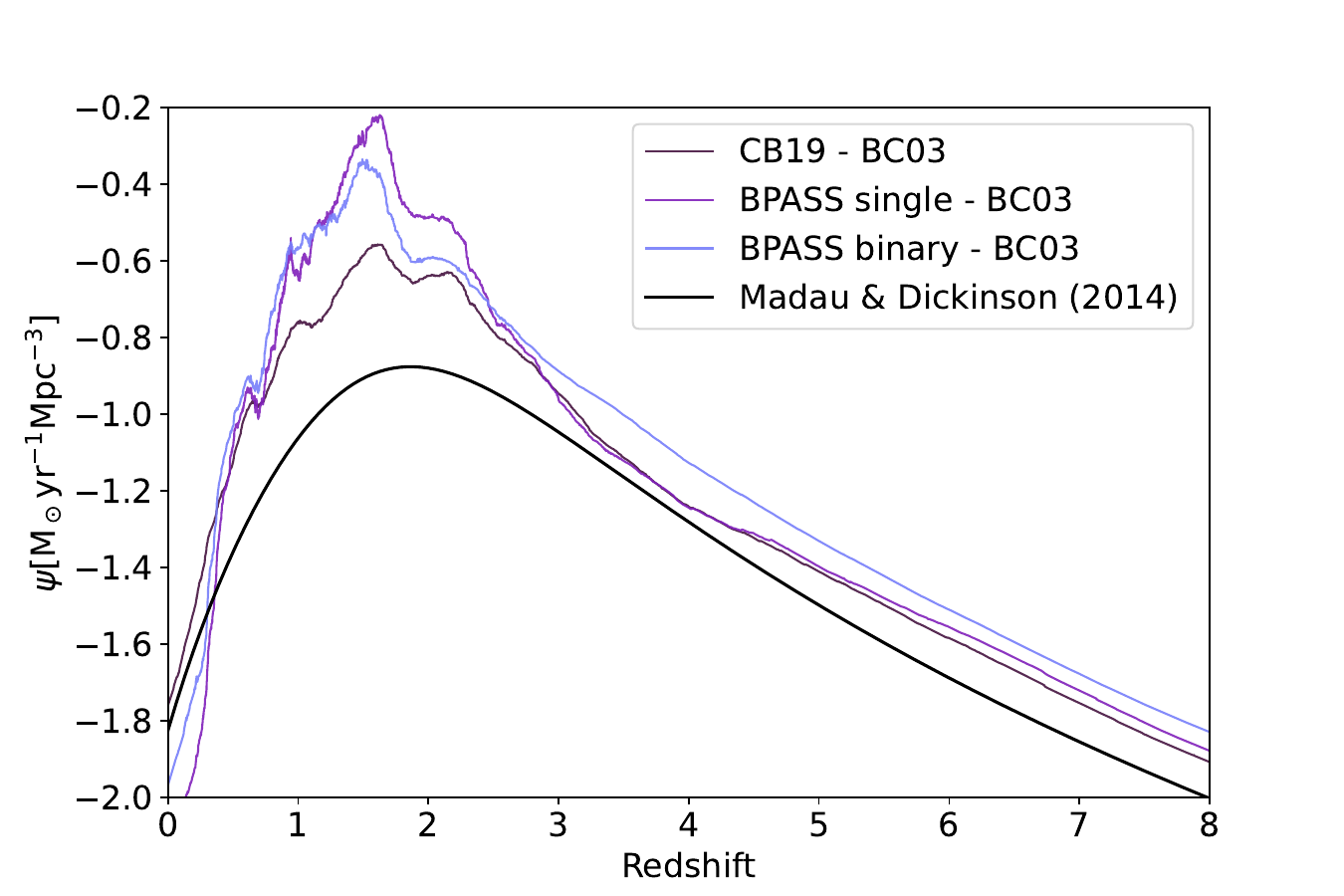}
    \caption{Reference curve of the cosmic SFRD from \citet{madau2014a} (solid black line) and the same curve after applying the SFR offsets inferred in this work.}
    \label{fig:sfrd}
\end{figure}

The SFR offsets and their relationship with the redshift can also be applied to a reference curve of the cosmic star formation rate density (SFRD) \citep[e.g.,][]{madau2014a}. In this way, we showcase the relevance of the obtained offsets to broader results. This is shown in Figure~\ref{fig:sfrd}.
We applied an offset rather than deriving a new curve, as our sample is not corrected for incompleteness.

To explore whether this effect is driven by the rest-frame wavelength coverage, we repeated the analysis after removing the four longest-wavelength bands, thereby mimicking the rest-frame coverage of high-redshift galaxies at lower redshifts. When we excluded those 4 bands, we limited the rest-frame coverage in such a way that for a $z=4$ galaxy, the longest band now is $\sim$ \SI{0.4}{\micro\metre} in the rest-frame. This is similar to how a $z=10$ galaxy is observed in the original runs. In this case, the offset peaks shift to a lower redshift, where even without the excluded bands, we still have rest-frame near-IR data. The plots in Figure~\ref{fig:seds} show the differences between the best-fit SEDs for three sets of galaxies at $z\sim0$, $z\sim2$, and $z\sim9$. Similar trends are found for other pairs of model families, as well as for the comparison between the two BPASS variants.

\subsection{Stellar population model preference}

As shown in Fig.~\ref{fig:chi-obs}, $\chi_r^2$ values indicate that SSP models generally provide acceptable fits. To assess whether any model is statistically preferred, we computed the Bayesian Information Criterion (BIC): $\mathrm{BIC} = \chi^2+k\log n$, where $k$ is the number of free parameters and $n$ the number of fitted photometric bands. Since $n$ and $k$ are identical for all models, differences in BIC reduce to $\Delta\mathrm{BIC}=\Delta\chi^2$. The difference in the BIC, $\Delta\mathrm{BIC}$, provides a statistical indication whether a model is favored over another.

\begin{figure*}[!htbp]
    \includegraphics[width=\columnwidth]{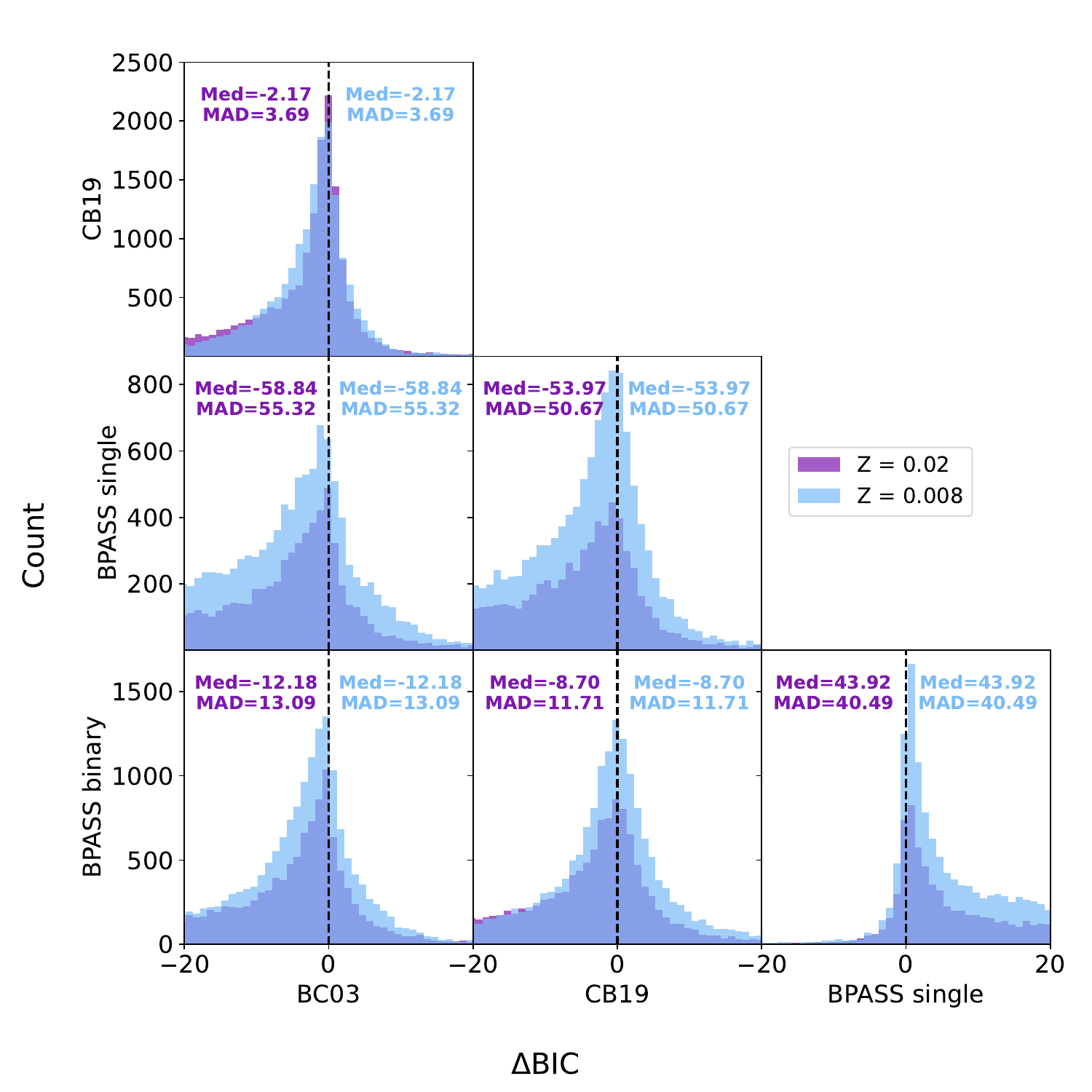}
    \includegraphics[width=\columnwidth]{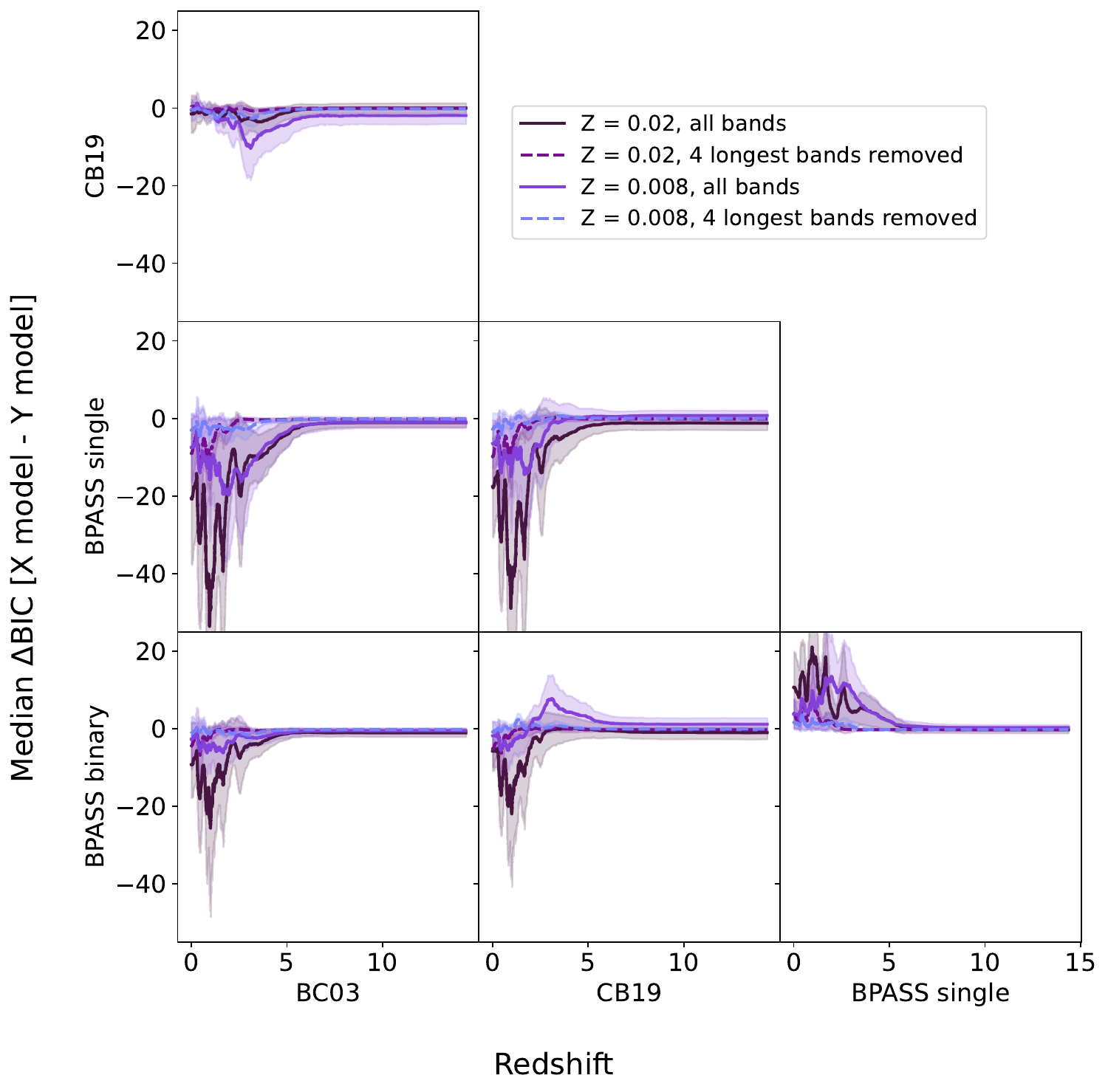}
    \caption{Left: Distribution of $\Delta\mathrm{BIC}$ on the fits of the observed galaxies. Right: Moving median of $\Delta\mathrm{BIC}$ with redshift for all combinations of models. The moving median was calculated using a moving window with a width of 1000 sources. The shaded regions represent the MAD.
    }
    \label{fig:bic}
\end{figure*}

Among the models, BC03 tends to yield the lowest overall $\chi^2$ values. The histograms in Figure~\ref{fig:bic} show the distributions of the $\Delta$BIC values for each comparison. We also show in Table~\ref{tab:bic} the percent of galaxies for which one model is strongly preferred over the other ($|\Delta\mathrm{BIC}| \geq 10$). Between the BC03 and CB19 models, most galaxies do not show a strong preference. In the comparison between the two BPASS variants, the binary-star version is strongly preferred for $\sim40-45\%$ of galaxies. Between model families, the BC03--CB19 family is generally strongly preferred more often. The fraction of galaxies exhibiting a strong preference for the BC03 model can reach up to $62\%$ between the BC03 and BPASS single-star models for $Z=0.02$.

\begin{table*}[]
\caption{Percent of galaxies showing a strong preference (|$\Delta$BIC| > 10) for either the model in the column name or the model in the row name for each combination of models.}
\label{tab:bic}
\centering
\begin{tabular}{c|c|cc|cc|cc}
$Z$                    & Model        & \multicolumn{2}{c|}{BC03} & \multicolumn{2}{c|}{CB19} & \multicolumn{2}{c}{BPASS single} \\ \hline
\multirow{3}{*}{0.008} & CB19         & 2.6\%        & 15.7\%     &              &            &                 &                \\
                       & BPASS single & 4.5\%        & 50.6\%     & 4.3\%        & 42.4\%     &                 &                \\
                       & BPASS binary & 4.2\%        & 26.0\%     & 8.8\%       & 18.5\%      & 40.8\%          & 1.3\%          \\ \hline
\multirow{3}{*}{0.02}  & CB19         & 2.0\%        & 13.0\%     &              &            &                 &                \\
                       & BPASS single & 2.5\%        & 61.8\%     & 2.2\%        & 58.8\%     &                 &                \\
                       & BPASS binary & 2.5\%        & 44.5\%     & 3.0\%        & 40.7\%     & 46.7\%          & 0.8\%          \\ \hline
\multicolumn{2}{c|}{strong preference for}            & row model   & column model  & row model   & column model  & row model      & column model     
\end{tabular}
\end{table*}

The redshift dependence of $\Delta\mathrm{BIC}$ follows a trend similar to that observed for the $M_\star$ and SFR offsets, decreasing toward higher redshift. At low redshift, there is a strong preference for the BC03 model, peaking around $z\sim1$. This preference weakens with increasing redshift and nearly vanishes at $z \gtrapprox 6$. The trend also becomes weaker when the four longest bands are excluded, again suggesting that the apparent differences arise from variations in rest-frame wavelength coverage rather than intrinsic model performance.

A similar behavior is found for both metallicities, though the amplitude of the variation is smaller for $Z=0.008$ than for $Z=0.02$. This result is broadly consistent with \cite{seille2024a}, who also find, using BIC comparisons, that BC03 models tend to provide better fits to photometric data.

\subsection{Effects on the main sequence}

As noted in Section~\ref{sec:intro}, reliable estimates of SFR and $M_\star$ are crucial for studying the SFMS. It is typically characterized by its slope, normalization, and scatter. These quantities, and their evolution with redshift, are key to understanding galaxy growth and the regulation of star formation. The normalization of the SFMS increases with redshift, reflecting the higher gas accretion rates in the early Universe \citep{santini2017a}. The slope constrains feedback mechanisms across mass scales \citep{popesso2019a, popesso2019b}, while the smaller intrinsic scatter ($\sigma \sim 0.25-0.4$~dex) implies that most star-forming galaxies follow a relatively similar SFH \citep[][and references therein]{santini2017a}.

\begin{figure}[!ht]
    \centering
    \includegraphics[width=\linewidth]{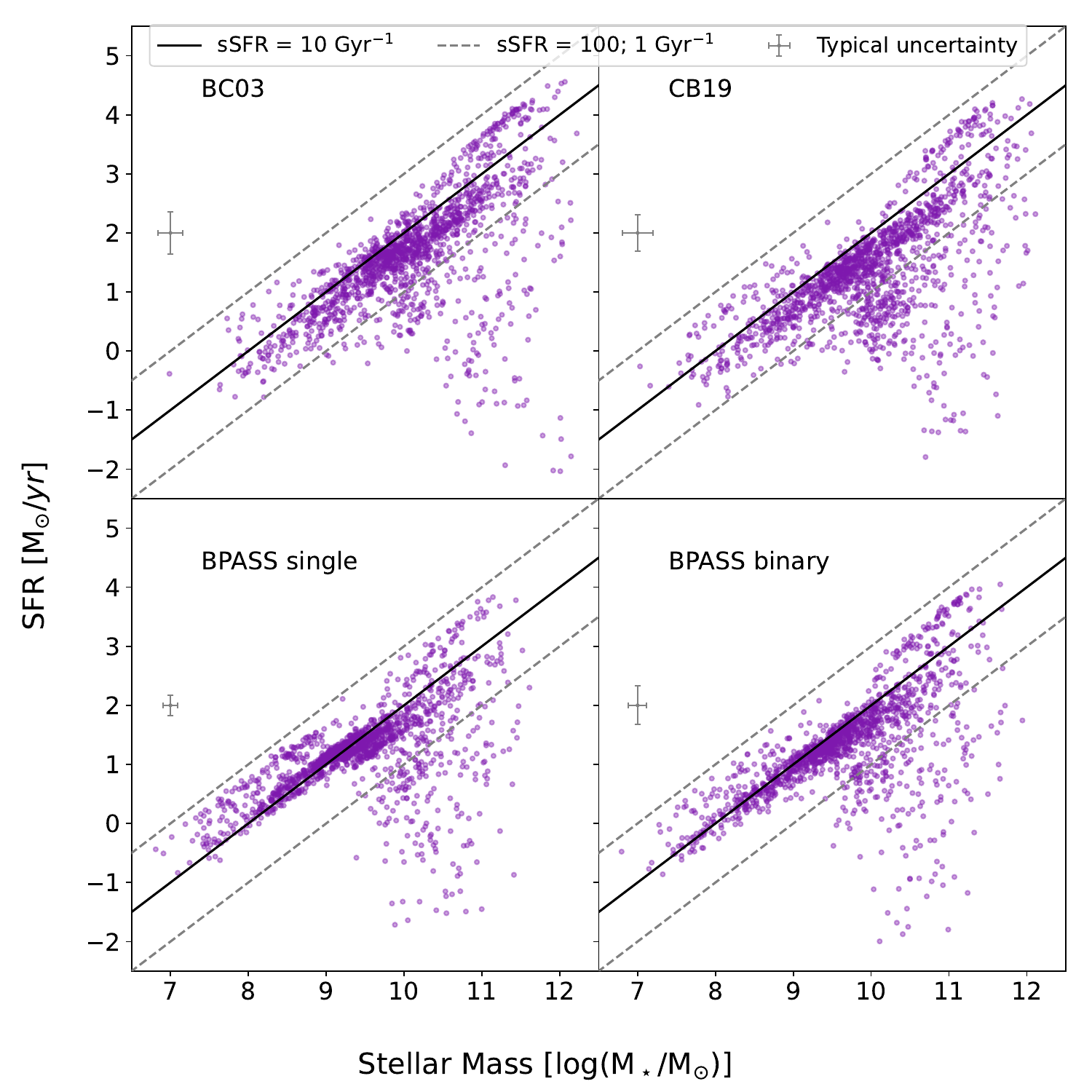}
    \caption{SFR - $M_\star$ plots for $z \sim 2$ with the results from all model runs on the observed fluxes. The solid black line represents a constant specific SFR (sSFR) of 10~Gyr$^{-1}$ and is not a fit to the data. The dashed lines show sSFR of 1~Gyr$^{-1}$ and 100~Gyr$^{-1}$.
    }
    \label{fig:MS}
\end{figure}

Following \cite{pacifici2023a}, we explore the impact of the assumed SSP model on the SFR-$M_\star$ relation for our sample at $1.8<z<2.2$, where the offsets between the inferred properties are largest. The corresponding plots are shown in Fig.~\ref{fig:MS}. Although a detailed SFMS fit is beyond the scope of this work, several qualitative trends emerge. Horizontal shifts reflect systematic differences in stellar mass estimates, while vertical displacements trace SFR offsets. For instance, the BC03 model yields fewer low-mass galaxies $(\log(M_\star/M_{\odot}) < 8)$ than the BPASS single-star model, consistent with the systematic offsets discussed earlier.

Overall, the measured differences in SFR and $M_\star$ are comparable with the intrinsic SFMS scatter reported in the literature, indicating that the choice of SSP model introduces non-negligible systematic uncertainties. Moreover, the BPASS models produce a larger number of galaxies with higher specific SFR, resulting in a slightly elevated upper envelope in the SFR-$M_\star$ plane.

\section{Summary and conclusions\label{sec:conclusion}}

Spectral energy distribution modeling is a widely used and efficient method for deriving the physical properties of galaxies, particularly at high redshift and in large samples, where data acquisition is both costly and time-consuming. In the era of the newest large (space and ground-based) observatories such as JWST, Euclid, ELT, Vera Rubin, etc., accurate and consistent SED modeling of extensive galaxy samples has become an essential task.

In this work, we investigate how the assumed SSP models affect the physical properties inferred from SED modeling. We compare several of the most widely used SSP models \citep{bruzual2003a, plat2019a, eldridge2017a} and quantify how model choice influences the derived SFR and $M_\star$ of galaxies. Our analysis is based on the catalog presented in \cite{merlin2024a}, which contains photometry for more than half a million sources observed in eight HST and eight JWST bands. We restrict the sample to galaxies with spectroscopically confirmed redshifts, resulting in a final dataset of 17,230 galaxies. The analysis is divided into two main parts:

\begin{enumerate}
    \item Fitting the observed fluxes: We fit all galaxies with each of the four SSP models (BC03, CB19, BPASS single, and BPASS binary) and compare the resulting physical properties -- SFR and $M_\star$.
    \item Generating and fitting synthetic photometry: In the second part, we generate synthetic galaxies based on the best-fit parameters from each of the four models. For each model, we produce a synthetic sample of 17,230 sources and simulate their photometric uncertainties. These synthetic datasets are then refitted using all four SPS models, allowing us to assess the internal consistency of each model and to compare the systematic differences due to the choice of SPS model with those arising from observational uncertainties.
\end{enumerate}

We find significant systematic offsets in the stellar masses derived from SED fitting, depending on the adopted SPS model. The BC03 and CB19 models yield consistent results with each other but show substantial offsets relative to the BPASS single and BPASS binary models. The inferred stellar masses differ by up to 0.56~dex between the BC03 and BPASS single models, which is larger than the propagated photometric uncertainties. For the SFR, the offsets reach up to 0.33~dex and are comparable to the uncertainties.

Examining the redshift dependence of these differences, we find a notable evolution in the offsets between model families (BC03/CB19 versus BPASS). We attribute this trend at least partially to the changing rest-frame wavelength coverage with redshift. Finally, repeating the analysis for two metallicities ($Z=0.02$ and $Z=0.008$) yields consistent results, indicating that our conclusions are robust against the adopted metallicity.

\begin{acknowledgements}
MB acknowledges support from the ANID BASAL project FB210003. This work was supported by the French government through the France 2030 investment plan managed by the National Research Agency (ANR), as part of the Initiative of Excellence of Université Côte d’Azur under reference number ANR-15-IDEX-01.
BS acknowledges support through an Erasmus Mundus Joint Master (EMJM) scholarship funded by the European Union in the framework of the Erasmus+, Erasmus Mundus Joint Master in Astrophysics and Space Science – MASS. Views and opinions expressed are, however, those of the author(s) only and do not necessarily reflect those of the European Union or the granting authority European Education and Culture Executive Agency (EACEA). Neither the European Union nor the granting authority can be held responsible for them.
PB acknowledges financial support through grant PRIN-MIUR 2020SKSTHZ and support from the Italian Space Agency (ASI) through contract ``Euclid - Phase E'', INAF Grants ``The Big-Data era of cluster lensing'' and ``Probing Dark Matter and Galaxy Formation in Galaxy Clusters through Strong Gravitational Lensing''.
PS acknowledges support from INAF RF2024 Large Grant "UNDUST: UNveiling the Dawn of the Universe with JWST".
\end{acknowledgements}

\bibliographystyle{aa}
\bibliography{SSPbib}

\clearpage
\onecolumn
\raggedbottom

\begin{appendix}

\section{2D histograms of offset variations with redshift}

\begin{center}
    \includegraphics[width=0.495\linewidth]{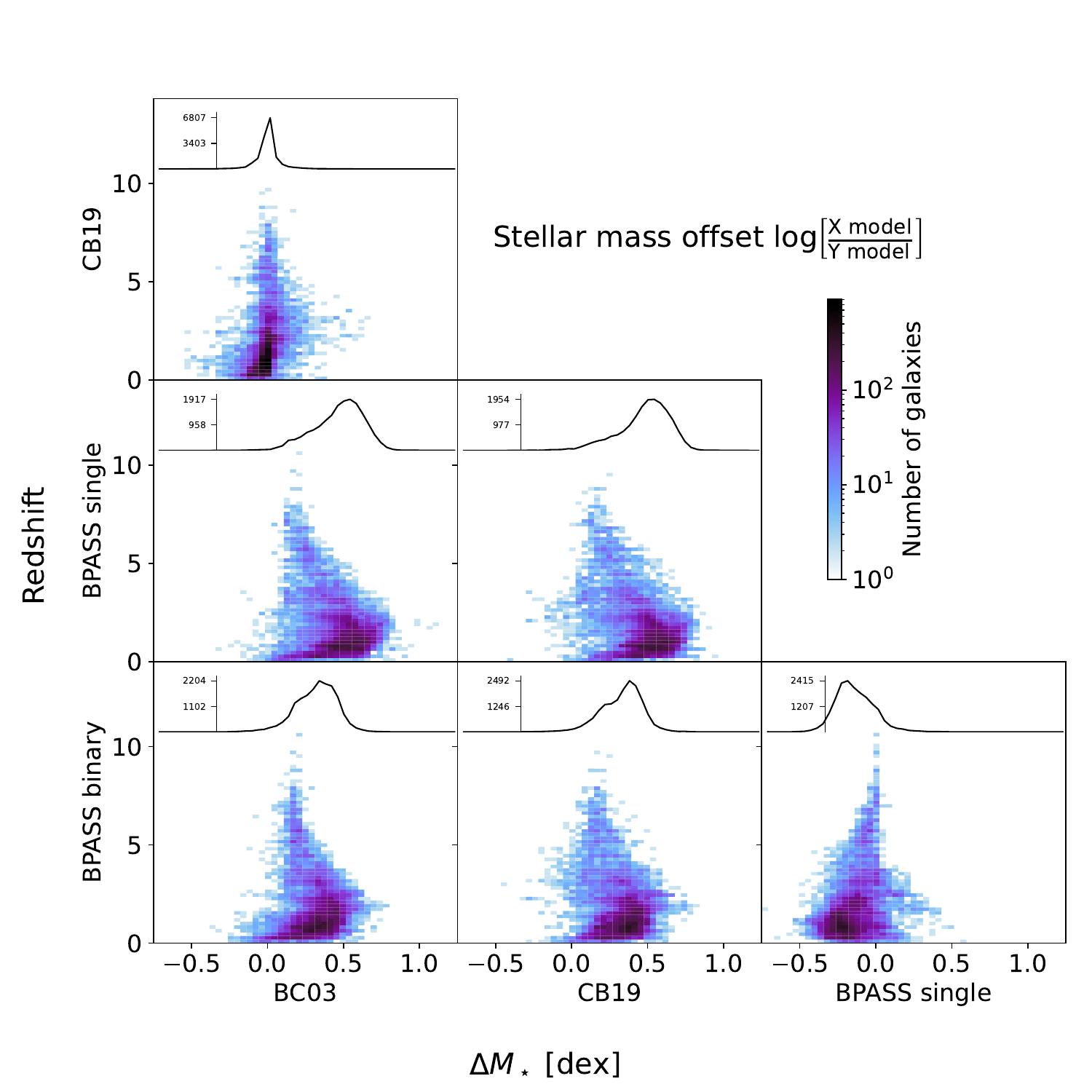}
     \includegraphics[width=0.495\linewidth]{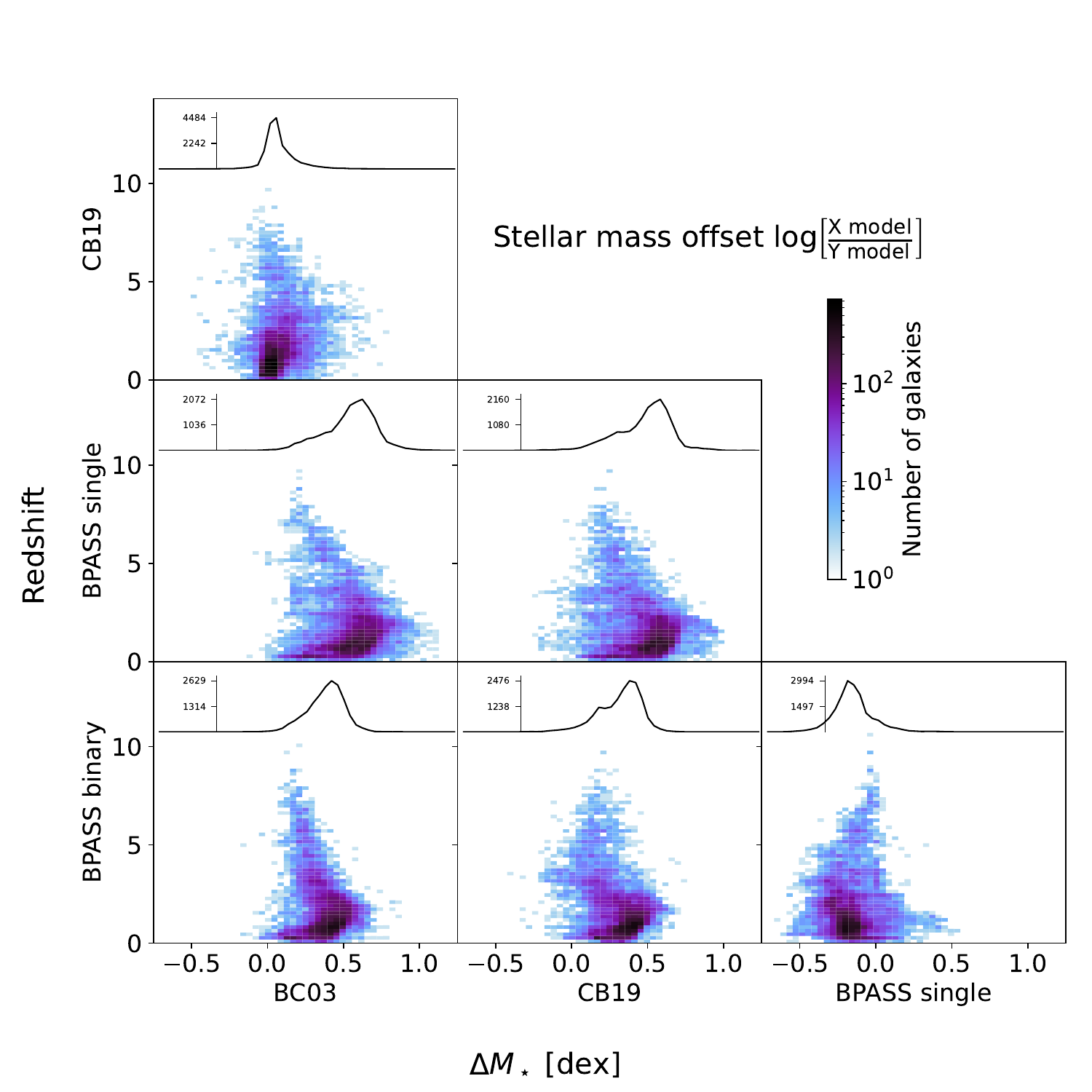}
    \captionof{figure}{A 2D histogram of the offsets with redshift. On top of each plot is a plot of the offset distribution integrated over the redshift. Left: $Z=0.02$. Right: $Z=0.008$}
    \label{fig:2dhist_mass}
\end{center}

\begin{center}
    \centering
    \includegraphics[width=0.495\linewidth]{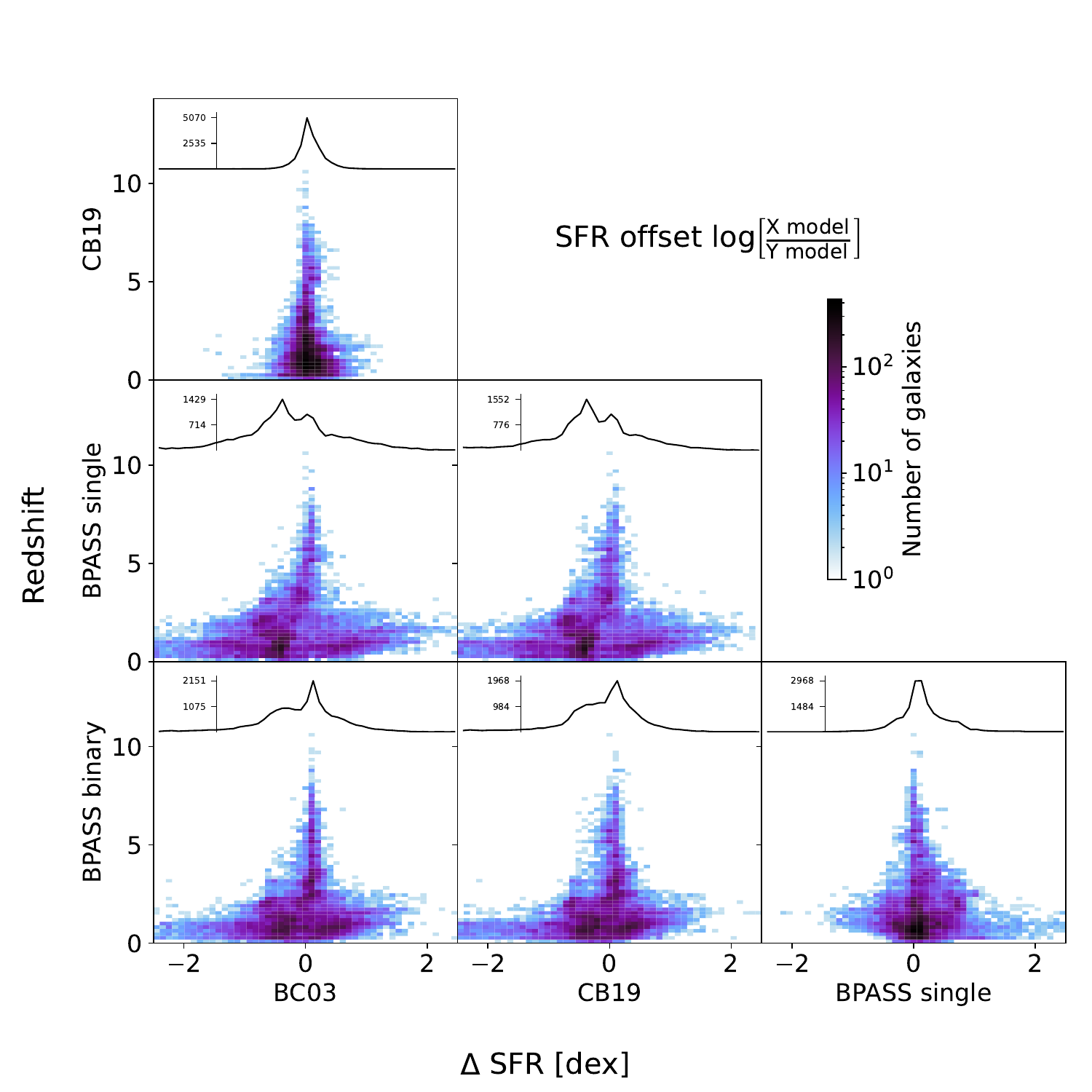}
    \includegraphics[width=0.495\linewidth]{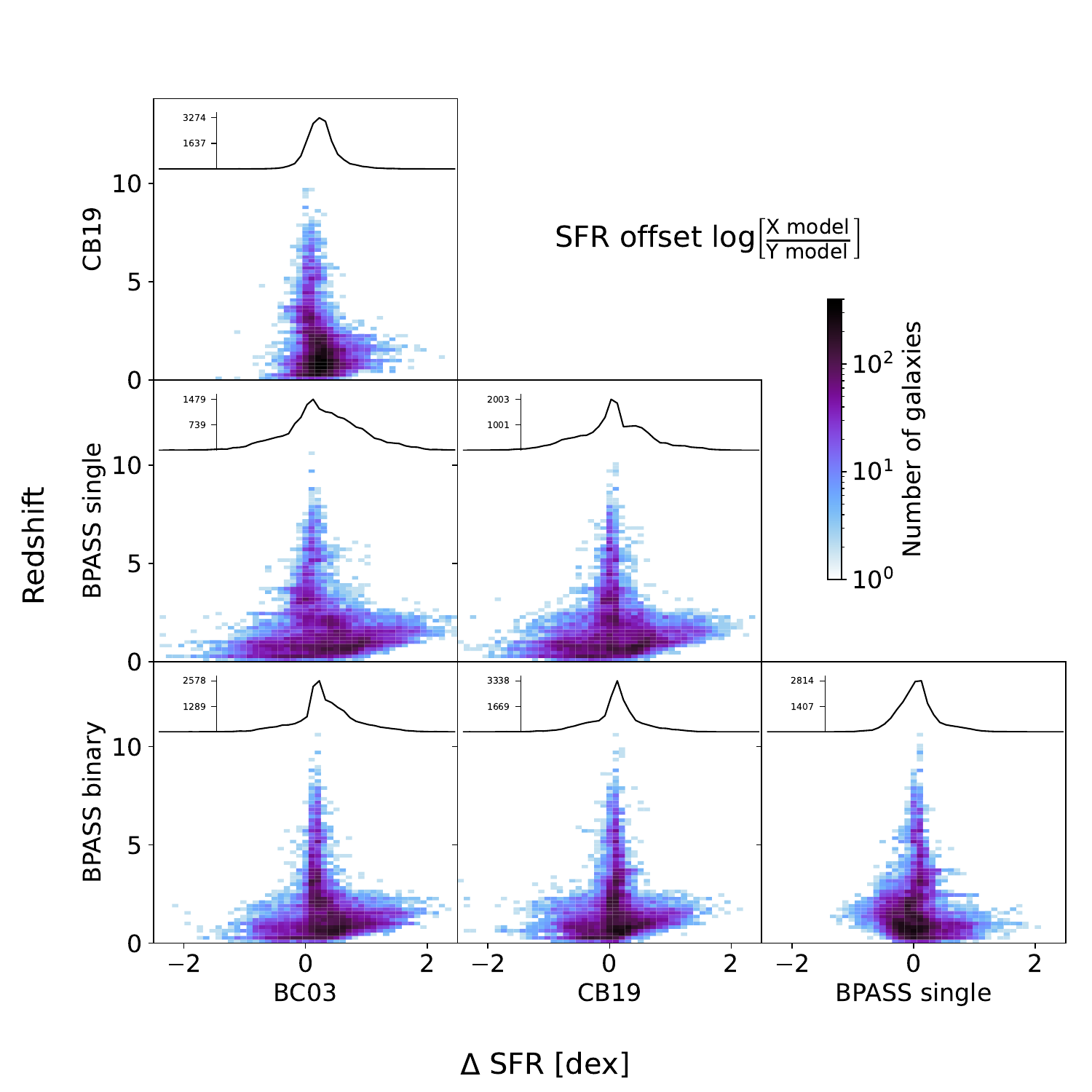}
    \captionof{figure}{Same as Figure~\ref{fig:2dhist_mass} but for SFR.
    }
    \label{fig:2dhist_sfr}
\end{center}

\newpage
\section{Additional direct comparison plots for $Z = 0.02$}

\begin{center}
    \includegraphics[width=0.475\linewidth]{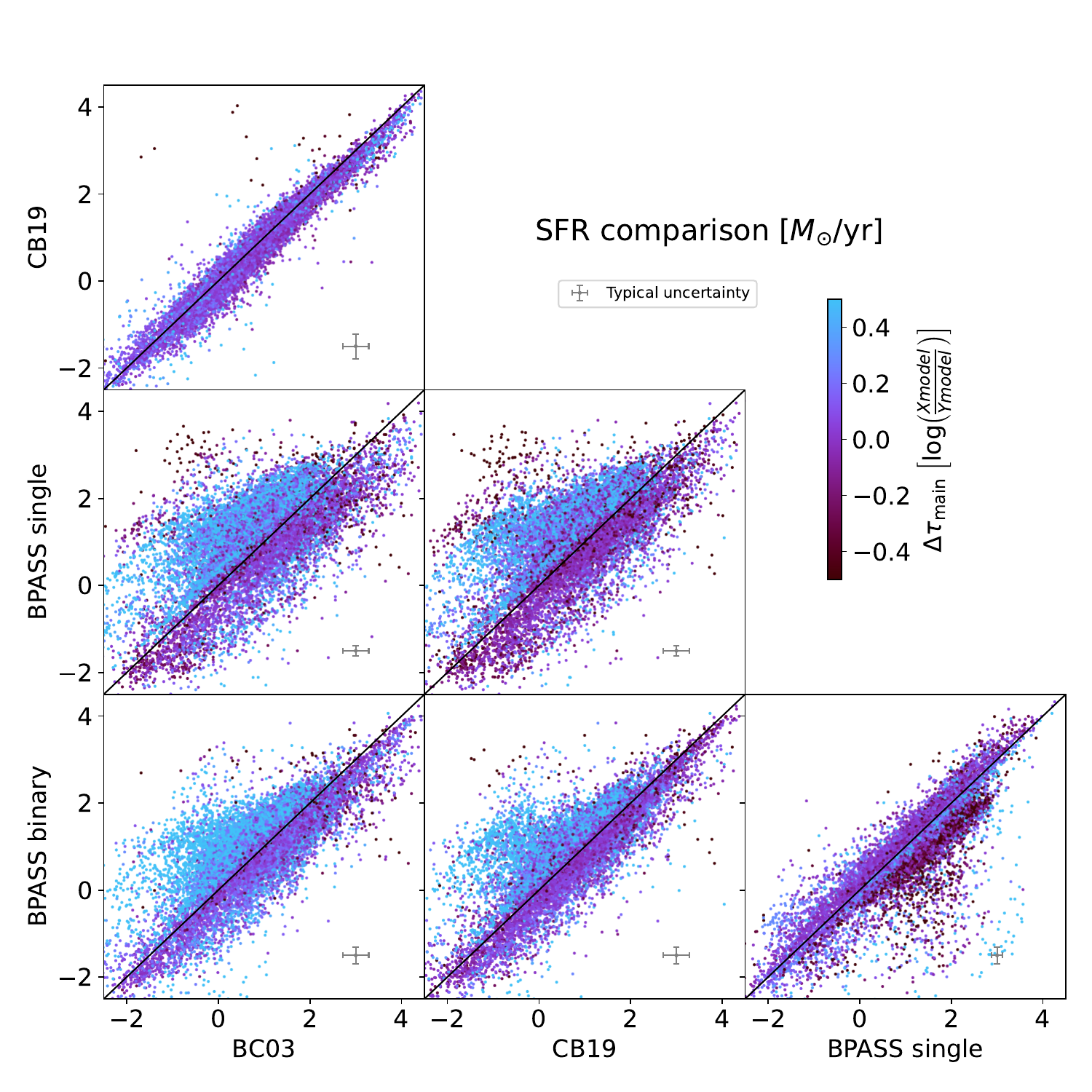}
    \includegraphics[width=0.475\linewidth]{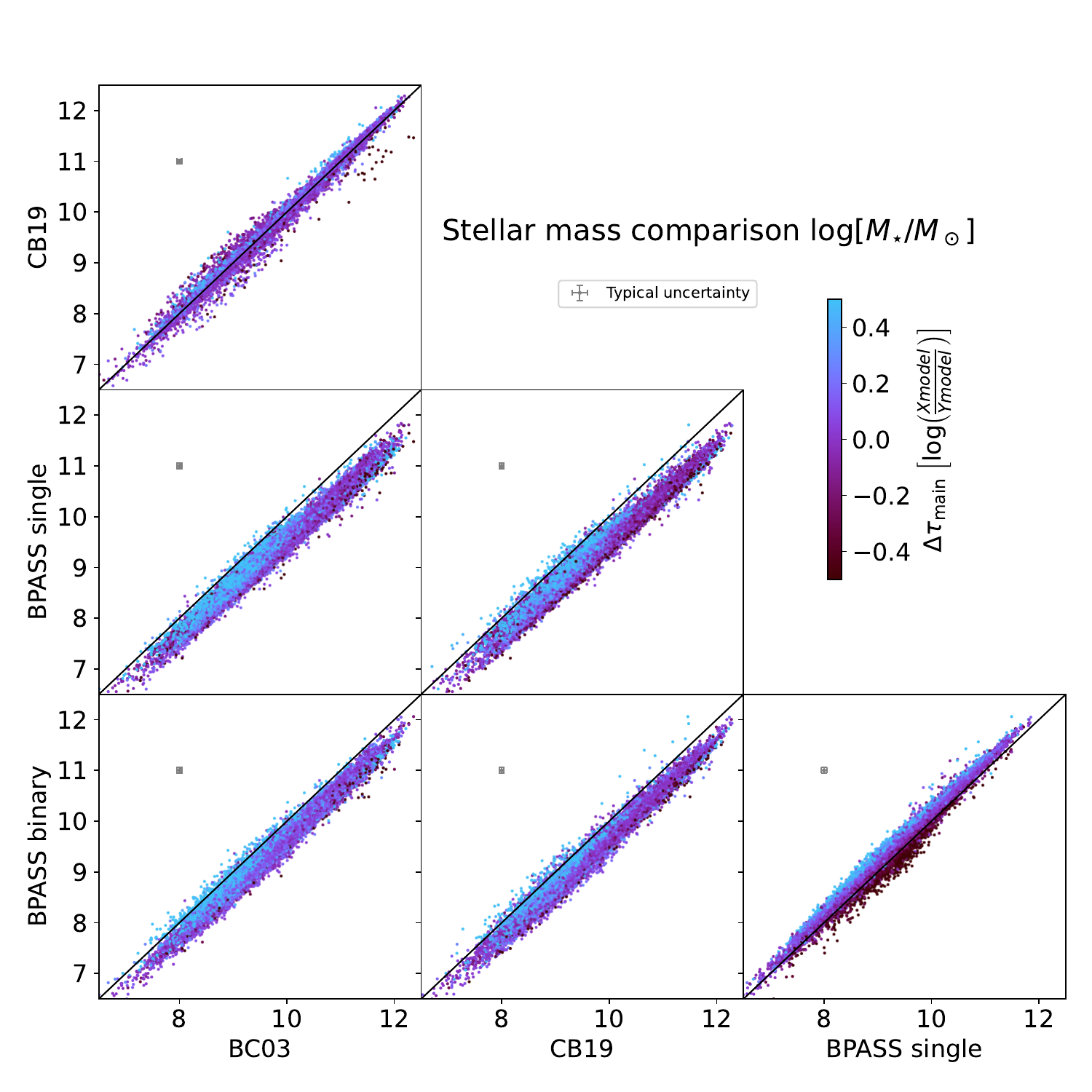}
    \captionof{figure}{Same as in Figure~\ref{fig:obs} but for $Z=0.02$.}
    \label{fig:obs_02}
\end{center}

\begin{center}
    \includegraphics[width=0.495\linewidth]{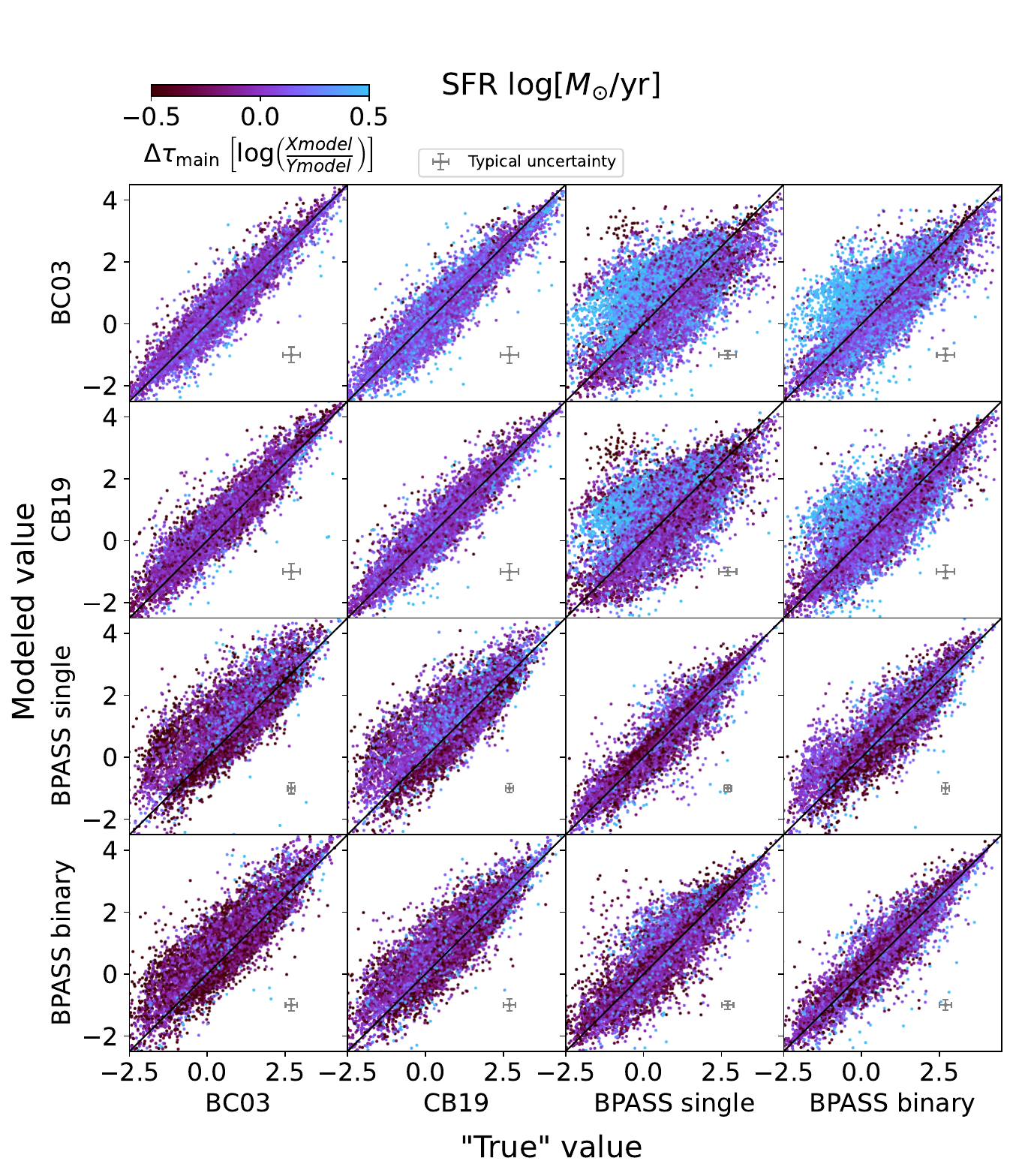}
    \includegraphics[width=0.495\linewidth]{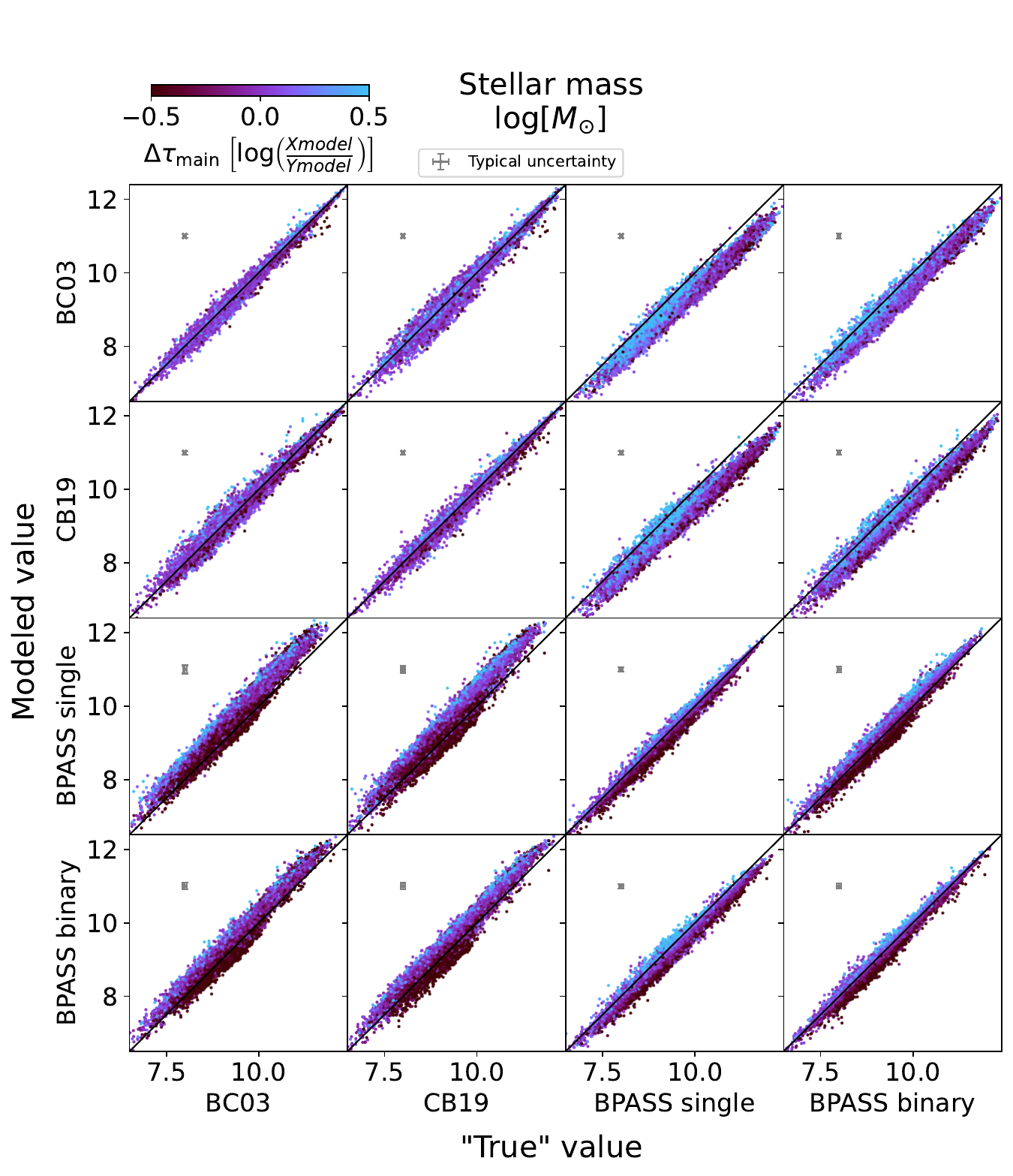}
    \captionof{figure}{Same as in Figure~\ref{fig:synth} but for $Z=0.02$.}
    \label{fig:synth_02}
\end{center}

\newpage
\section{CIGALE modules parameters}
\begin{center}
\captionof{table}{Modules and their corresponding parameters used for this work to build a grid of 289 536 models per redshift for every set of stellar population models.}
\label{tab:params}
\begin{tabular}{l|lll}
\hline \hline
                                                                                          & Module                  & Parameter       & Value                                                                 \\ \hline
\multirow{09}{*}{\begin{tabular}[c]{@{}l@{}}Common\\ modules\end{tabular}}                & sfhdelayedbq            & tau\_main       & from 1 Myrs to 13 500 Myrs with a 500 Myrs step ; 30 000                                  \\
                                                                                          &                         & age\_main       & from 100 Myrs to 13 600 Myrs with a 100 Myrs step\\
                                                                                          &                         & age\_bq [Myrs]        & 1, 3, 5, 7, 10, 15, 20, 30, 40, 60, 80, 100, 150                                                  \\
                                                                                          &                         & r\_sfr          & 0.01, 0.05, 0.1, 0.5, 1, 5, 10, 100 \\ \cline{2-4} 
                                                                                          & nebular                 & logU            & -2.0                                                                  \\
                                                                                          &                         & zgas            & 0.02, 0.008 \tablefootmark{*}                                                           \\ \cline{2-4} 
                                                                                          & dustatt\_modified\_CF00 & Av\_ISM         & 24 linearly spaced values from 0.0 to 3.0 mag                         \\
                                                                                          &                         & mu              & 0.2, 0.3, 0.4, 0.5                \\ \cline{2-4} 
                                                                                          & redshifting             & redshift        & —                                                                     \\ \hline
\multirow{10}{*}{\begin{tabular}[c]{@{}l@{}}Stellar \\ population\\ modules\end{tabular}} & bc03                    & imf             & 0 (Salpeter)                                                          \\
                                                                                          &                         & metallicity     & 0.02, 0.008 \tablefootmark{*}                                                                \\
                                                                                          &                         & separation\_age \tablefootmark{a} [Myrs] & 10                                                                    \\ \cline{2-4} 
                                                                                          & cb19                    & imf             & 0 (Salpeter)                                                          \\
                                                                                          &                         & metallicity     & 0.02, 0.008 \tablefootmark{*}                                                              \\
                                                                                          &                         & upper \tablefootmark{b} [Myrs]       & 100                                                                   \\
                                                                                          &                         & separation\_age \tablefootmark{a} [Myrs]& 10                                                                    \\ \cline{2-4} 
                                                                                          & bpass                   & imf             & 2 (Salpeter)                                                          \\
                                                                                          &                         & metallicity     & 0.02, 0.008 \tablefootmark{*}                                                                \\
                                                                                          &                         & binary          & 0, 1 \tablefootmark{*}                                                                \\
                                                                                          &                         & separation\_age \tablefootmark{a} [Myrs]& 10      \\ \hline \hline                                                             
\end{tabular}
\tablefoot{Parameters that are not shown in the table were left at their default values. 
\tablefoottext{*} {Kept fixed for each individual run, but all given values were fit in separate runs.}
\tablefoottext{a} {The separation between the old and young stellar population.} 
\tablefoottext{b} {The upper-mass cutoff of the IMF. This is a parameter only for the CB19 model and is left to the default value of 100~Myrs to match the assumed value in the other models.}}
\end{center}

\section{Example SEDs}

\begin{center}
    \includegraphics[width=0.719\linewidth]{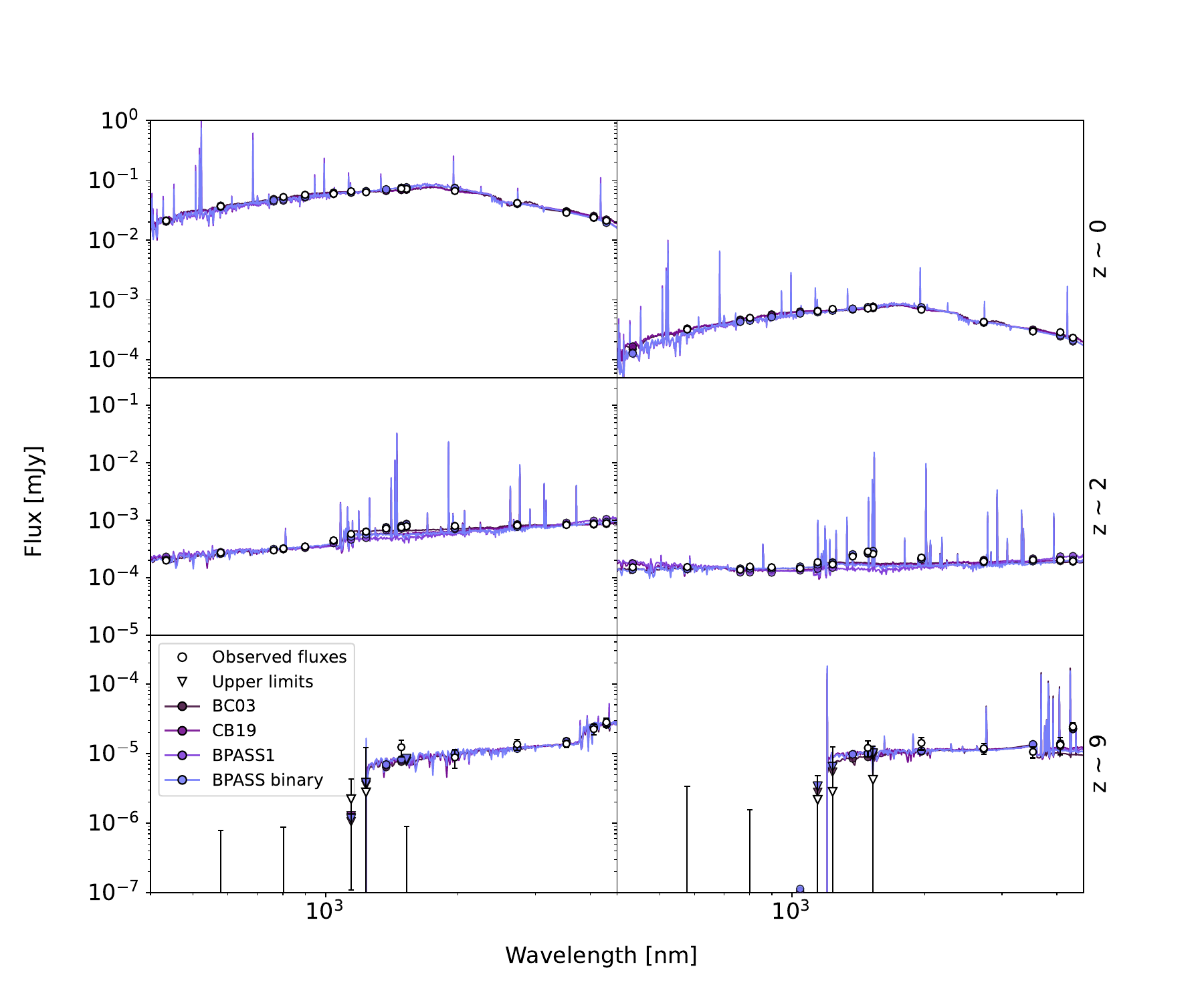}
    \captionof{figure}{Examples of the best fit SEDs obtained with each of the stellar population models for galaxies exhibiting the typical stellar mass offsets between models at three different redshifts: $z\sim0$ (top), $z\sim2$ (middle), $z\sim9$ (bottom).}
    \label{fig:seds}
\end{center}

\end{appendix}
\end{document}